\documentclass[12pt]{article} 
\usepackage{multicol} 
\usepackage{comment}
\usepackage[a4paper, top=0.7in, bottom=1in, left=0.7in, right=0.7in]{geometry} 
\usepackage[table]{xcolor}
\usepackage{pgfplots}
\usepackage{pdfpages}
\usepackage[]{easymcm}  
\usepackage{indentfirst}
\usepackage{float}
\usepackage{svg}
\problem{E}  
\usepackage{tikz}
\usetikzlibrary{arrows}
\usepackage{caption}
\usepackage[affil-it]{authblk}
\usepackage{graphicx} 
\usepackage{subcaption}
\usepackage[table]{xcolor}
\usepackage{graphicx}

\begin{document}

\marginpar{\tiny Authors \\ ranked \\ by last \\ names.}

\begin{center}
\fontsize{20}{12}\selectfont\textbf{Re-imagining the Future of Forest Management — An Age-Dependent Approach towards Harvesting}  
\end{center}
\vspace{4mm} 

\begin{center}
Shuyang Bian, Yuanyuan Xie, Flora Zhang \\
College of Arts and Sciences, Emory University\\
Primary Contacts: simon.bian@emory.edu, yuanyuan.xie@emory.edu
\end{center}

\vspace{5mm}
\begin{multicols}{2}

\section{Abstract}
    
Facing the drastic climate changes, current strategies for enhancing carbon dioxide stocks need to be thoroughly honed. While technological developments improve the quality of human life, they also severely damage the global carbon cycle with $CO_2$ emissions. Sole preservation is not a panacea, for it may well pack the woods, resulting in fire and ultimately carbon emissions. We attempt to search for a well-designed model of forests' carbon storage management, such that harvested carbon may yield economic return without compromising the delicate balance with nature.

To address the problem, we first built a carbon sequestration growth model driven by \textbf{growth rate dependency (GRDM)}. We abstracted the carbon cycling system into the process of photosynthesis, the humidity fluctuation, and the original storage of carbon in the trees. In the photosynthesis model, we considered various factors, including transition rate of absorption and organic matter production. In the humidity influence model, we defined the range of humidity that assisted plant growth and approximated the volume of carbon storage. The carbon storage in each tree is denoted based on the summation of per unit volume of carbon in each branch. Further, noticing the age-based differences in species ability to store carbon, we simulated the carbon harvest based on natural selection.

We also designed an \textbf{Economic Return Evaluation Model (EREM)} to estimate the optimal distribution of trees in the forest based on the utility function. Maximizing the utility brought by the amount of carbon storage, we derived the equation for profit optimization with the constraints of total economic expenses allowed.

To assess its performance, we took an \textbf{object-oriented approach}, simulated an $8\times8$ ideal forest by placing instances of trees and plotted a time-dependent forest composition graph. With a time interval of 100 years, we recognize that the forest management strategies should be adjusted with a combination of three approaches: age-dependent, height-dependent, and density-dependent approach. Model-wise speaking, we find that selecting for trees' age would typically yield higher trees, a subsequent reduction in density, and smaller wildfires. Parameters adjustment show that when cutting trees whose ages are within $10\%$ of the age at which wood loses its carbon sequestration ability, we will not compromise environmental protection for the sake of economic benefits. After proper normalization of climate and economic data, we also make predictions for 169 worldwide forest-covered countries. Our model further suggests high sensitivity and robustness with a similar trend of overall utility when environmental aridity or proportion of harvested woods are varied. 

Finally, we apply the model to Georgia temperate deciduous forest, and we evaluate the carbon storage ability to adjust the Red Spruce based on available biological literature research. We recognize that while the model is preliminary in its failure to identify a diverse array of variables, it has encapsulated key features of idealized forests.

     \vspace{5pt}
     \textbf{Keywords}: Carbon Cycle, Utility Model, Time-dependent Analysis, Management Optimization.

\section{Introduction}
\subsection{Problem Background}
As organisms of the primary tropic level, green plants - with trees in forests contributing the most - fixate carbon dioxide by photosynthesis, thereby contributing to the circulation of global carbon flow and acting as a carbon sink. $^{[1]}$ Human activities and logging, nevertheless, serve as a carbon source that contributes to the increase of global $CO_2$. $^{[2]}$

While it seems natural that minimizing forest use to be feasible in preserving natural environments, studies reveal the possibility of striking a balance between economic benefits of carbon sequestration and maintaining forest vivacity. $^{[3]}$ 

Furthermore, inter-species competition may arise when succession occurs, meaning that without proper forest management plans, prior fundamental species are replaced and converted into detritus, acting as a carbon source.$^{[1]}$ Environmental conditions including forest fires may also occur under high species density$^{[4]}$ more frequently, hence releasing carbon dioxide.

\subsection{Problem Restatement}
We need to create a carbon sequestration model to estimate the amount of carbon dioxide a forest will store and absorb over time. The model identifies the carbon sequestrating management strategy to maximize forest's competence. After that, we develop a decision model to assist forest managers in determining the most efficient usage of a forest. The model develops a forest management strategy that considers many ways forests are valued including carbon sequestration, biodiversity aspects, and recreational uses, etc.

Considering the spectrum of management strategies, we need to design the decision-making model concerning tree types, geography, and lifespan, etc. We will test the model with various real-life forests to adjust the error term of the model. With the advanced model, we predict the carbon dioxide a forest produces and absorbs over the next 100 years. With the optimum management plan and the prediction model, we suggest possible transitioning strategies for all forest users to add on to existing practices. Finally, as the forest is more and more open to the public, we support the statement of limiting the trees to optimal amount with the model.
\subsection{Scenario-based Methodologies}
\begin{enumerate}[\bfseries 1.]
    \item Regarding net carbon dioxide absorption, we use ordinary differential equations to model the growth rate dependency(GRDM). As known$^{[1]}$, growth rate positively correlate with carbon retention. We consider three chronology-dependent aspects, including environmental temperature, overall humidity, and biological characteristics. 
    \item In light of the gain of net carbon dioxide absorption from the previous model, we then model rival factors, including human use of plants. Using the limitation of budget model (LBM), we consider aspects including human resource investment in forest management and commodity values of carbon sequestration.
\end{enumerate}

Finally, a utility function (UF) is established to balance the gains and losses of both models. And different approaches were examined using the mathematical model.

\section{Carbon Sequestration Models (CSM)}
\subsection{Assumptions}
\begin{enumerate}[\bfseries 1.]
\item Concerning the chemicals involved in this model, we assume that no atoms other than Carbon are discussed for this mathematical model.
\item  Concerning the biological systems involved in global carbon cycles, it is suggested that

(1) Unless otherwise amended, growth of trees are affected by no other factors than age, volume, temperature and moisture availability.

(2) No other biomass other than trees in forests should be discussed. Further, trees in the forest rigidly do not expire or their expiration rates are also less than their inspiration of $CO_2$.

(3) Human activities involve only logging and bringing carbon dioxide to the atmosphere, and its only economically favorable outcomes are the sequestered carbon from woods.

Other specific assumption, if necessary, will be mentioned and illustrated while we’re building the models.
\item Concerning the mathematical model we use, we consider that:

(1) The tree can be modeled in a cylinder and whose volume would change in a predictable way;

(2) Once the trees were cut for carbon sequestration, the harvested carbon would no longer emit or absorb any other chemical, and also that once cut, the trees were considered dead. Seedlings would go from the exact location at which sawing happen.

\end{enumerate}

\subsection{Notations}
The primary notations used in this paper are listed in the table in the last few pages.

If additional notations are necessary, they will be explained and illustrated while building models.

\subsection{Carbon Sequestration Model: Growth Rate Dependency (GRDM)}
The variation of carbon in forests over time is written as the total of the carbon stored and the carbon used for volume growth: \\
\begin{equation}
    \frac{dC_{tot}}{dt}=F(H(t,T),h,rad,s,A)+ P(C_a,T,h,rad, \psi) + \frac{dS_c}{dt}
\end{equation}

Here, F is the gross carbon storage rate due to Humidity($H$), height of the tree($h$), radius to the tree trunk($r$), number of trees harvested($s$), and the age of the tree($A$). P denote the carbon assimilation due to gross photosynthesis eliminating the usage of foliage respiration. However, compare to the total carbon storage, the respiration of plants is almost negligible under our assumption. 
\begin{equation}
    H(t,T)=H (sin(\omega t),T(t)=\kappa cos(B \omega t)+D)
\end{equation}

Among all species, there are non-resistant $\beta$ species which takes $\frac{dC_{tot}}{dt} \propto H$, the rest $1-\beta$ drought resistant thrives under low humidity such that $\frac{dC_{tot}}{dt} \propto \frac{1}{H}$. Notice that such periodic relationship only works when $T \leq 451 ^{\circ}F $, that is, when the temperature has not exceeded the point to set wood on fire. As shown below, temperature is also a function of time t. m represents the mass of the tree at time $t$, while $\chi$, $\sigma$, $\gamma$, $\kappa$ are constant parameters, and $\epsilon$ is the total error term. 
\begin{equation}
    H(t,T)=e^{-\frac{\chi}{2m}t}(\gamma sin(\omega t))+  e^{-\frac{\chi}{2m}t} B cos(\omega t)) + \epsilon
\end{equation}
\begin{equation}
    \text{  with }\omega = \frac {\sqrt{\chi^2-4m\kappa}}{2m}
\end{equation}
\begin{equation}
\text{ and }m=\rho \text{V}=\rho \left(\pi {{rad}}_{1}^{2}h+\frac{2}{3}\pi {rad}_{2}^{3}\right)
\end{equation}

The above eq.3 would give a graph with oscillation behavior that tends to a set humidity $H_{0}$ asymptotically with the setting of constant $\omega$ (eq.4) and mass at time t $m$ (eq.5).

\begin{figure}[H]
    \centering
    \includegraphics[width=9cm]{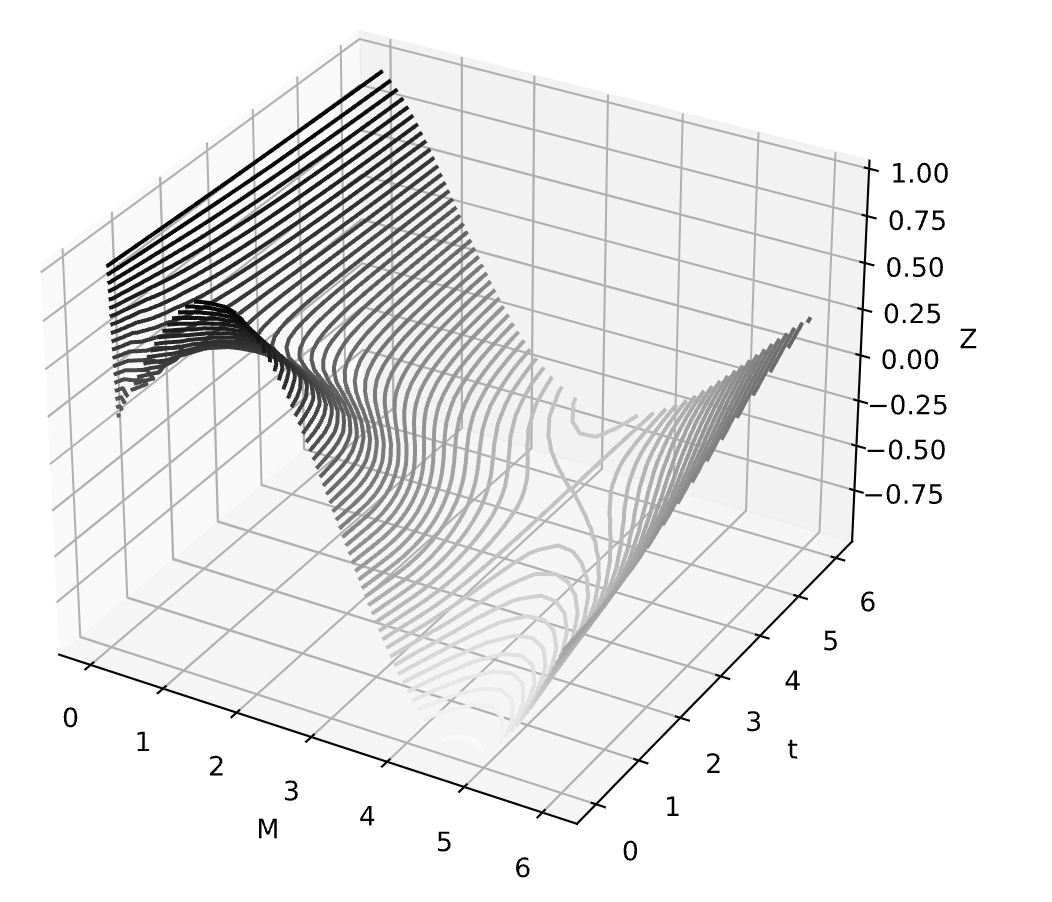}
    \caption{Three-dimensional visualization of formula (3)}
    \label{fig:my_label}
\end{figure}

\begin{figure}[H]
  \centering
  \includegraphics[width=9cm]{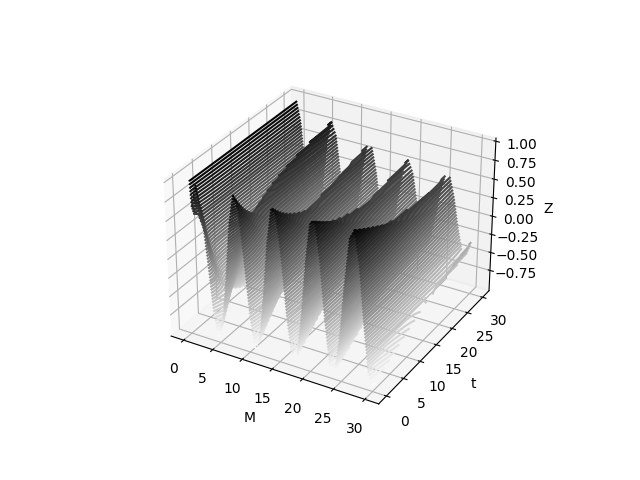}
  \caption{Three dimension of formula (3) asymptotic behavior}
\end{figure}

\begin{equation}
\begin{aligned}
    S_{c} &= S\frac{\lambda }{\left(1+\frac{dV}{dt}\right)}, \\
    \text{where } V &\approx \pi \text{ rad}^{2}h.
\end{aligned}
\end{equation}

Approximating the volume of tree trunks with the shape of cylinder, and the volume of canopy with a half of a sphere, we have the storage of carbon to the relation of the volume as eq.6 above.
\begin{equation}
\frac{dC_{tot}}{dt}=\frac{1}{\mathrm{s}}
\end{equation}

Harvesting trees would bring the carbon storage within each tree directly out while remain nearly neglectable damage to the environment. Such linear relationship is denoted by eq.7 above.

P is modeling photosynthesis considering the influential factors of atmospheric $CO_{2}$, the growth rate of the tree $\frac{h\mathrm{'}-h}{h\mathrm{'}}$, the temperature of the growing environment $(T-T_{opt})$, and water potential. 
\begin{equation}
P\alpha (1-\exp C_{a})
\end{equation}

Photosynthesis is positively related to 1 minus the exponential of $CO_{2}$ concentration in the atmosphere $C_a$.
\begin{equation}
P\alpha \left(\Psi +\frac{2}{3}\pi rad^{3}h\right)
\end{equation}

The solar radiation absorbed is positively influenced by the depth of each canopy of tree, photosynthesis tends to decrease with declining water potential due to the height of the trees. 
\begin{equation}
P\alpha (1-\left(\mathrm{T}-\mathrm{T}_{\mathrm{opt}}\right)^{2})
\end{equation}

Finally, in the assumed test environment of rain forests, photosynthesis’s dependence on temperature is approximately a concave down parabola. Hence, $P$ increase with the water potential minus the temperature difference to the optimal temperature($T_{opt}$) squared.

Combining relations, we have from (8), (9), and (10), we have:

\begin{equation}
    P = (1-exp(C_a)) (\psi+2/3 \pi rad^3 h) (1-(T-T_{opt} )^2) 
\end{equation}

There are four defined stages for the age of trees: birth to sprout, sprout to youth, youth to mature, and mature to death. For all four stages, we have $\frac{dC_{tot}}{dt} \propto -\ln(A)$ for each tree.$\mu$ represents the constant that compensates for the errors.
\\Thus, we get 

\begin{equation}
    \frac{dC_{tot}}{dt}=-\ln(A)+\mu
\end{equation}

\subsection{Utility function based - Economic Return Evaluation Model (EREM)}

Utility function  $U(C_{tot},E)$ is denoted by:
\begin{align}
\nonumber
U(C_{\text{tot}},E) = &\theta \ln\left(\frac{1}{C_{\text{tot}}}\right) \\
&+ (1-\theta)\ln E + \tau \text{Simp} \ln(452-T)
\end{align}

Where $\theta$ represents the number of drought resistant Keystone Species and $(1-\theta)$ denotes the rest of the species which are non-resistant.The constant $\tau$ weights the influence of drought resistant species to utility.

With the restrain of $P_1C_{tot}+P_2E=I$, we apply the Lagrangian Method:

\begin{equation}
\nonumber
\begin{aligned}
    \max_{C_{tot},E,\lambda} U(C_{tot},E) &= \theta \ln\left(\frac{1}{C_{tot}}\right) \\
    &\quad + (1-\theta)\ln(E) + \tau \ln(T-452) \\
    &\quad - \lambda(P_1 \cdot C_{tot} + P_2 \cdot E - I)
\end{aligned}
\end{equation}

Here, the inverse Simpson index has no effect on utility when temperature T goes over $451 ^{\circ}F$, the burning point of wood, and does not make sense when the temperature goes higher than the limit. That is, we also need the condition of $T \geq 451 ^{\circ}F$.

We take the First Order Condition (F.O.C) as below:
\begin{equation}
\nonumber
    \frac{\partial U}{\partial C_{tot}}=\theta C_{tot} - \lambda P_1=0
    \end{equation}
\begin{equation}
\nonumber
    \frac{\partial U}{\partial E}=\frac{1-\theta}{E}-\lambda P_2=0
\end{equation}
\begin{equation}
\nonumber
\frac{\partial U}{\partial \lambda}=P_1C_{tot} + P_2E - I = 0
\end{equation}

from the first two equation, we have: 
\begin{equation}
\nonumber
P_1=\frac{\theta C_{tot} P_2 E}{1-\theta}
\end{equation}

then we rewrite our equation of $\frac{dU}{d\lambda}$ as:

\begin{equation}
\nonumber
    \frac{\theta C_{tot} P_2 E}{1-\theta} + P_2 E=I
\end{equation}

Therefore, \[ 
P_1 = \frac{I}{(1-\theta)C_{\text{tot}}E} 
\]
 and \[ 
P_2 = \frac{I}{\theta C_{\text{tot}}^2 E^2} 
\]

\section{Decision Model}
\subsection{Baseline Approach - A one-hundred year simulation}

Before taking on any forest management plans, we attempt to determine the projected forest growth without human intervention. We devise a Java algorithm (see Attachment) that takes into account different characteristics of trees. As shown in the UML diagram below, trees were considered as a class and were inherited into the forest class. Different properties of trees, including age, water-sensitivity, growth rate, and aging, are considered. These tree instances are then placed in an $8\times8$ 2-D array that would mimic a typical forest distribution. Based on available literature and assumption, the dryness of the environment would occur in a time-dependent, trigonometric way; and under a given set threshold, this would result in all trees decimated in a $4\times4$ grid.
Data is obtained with randomized initial status. Carbon dioxide absorption was calculated grid by grid. Colorization of data is achieved through normalized scaling as follows. 

\includegraphics[width=\columnwidth]{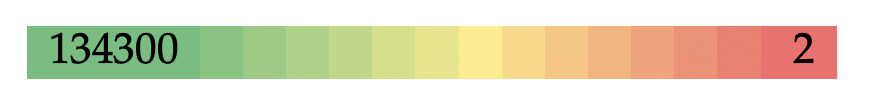}

The below represents a time-based change of carbon dioxide with random forest fire, tree death, and tree growth over a period of 100 years. Two colored grid cells at the start of the simulated experiment and the end of it were included above the trendline of total carbon absorption. It is clear that after the disturbances by fire, forests need restoration. Evidently, with zero human intervention, forest fires would occur and yield no particular benefits. \\

\includegraphics[scale=0.5]{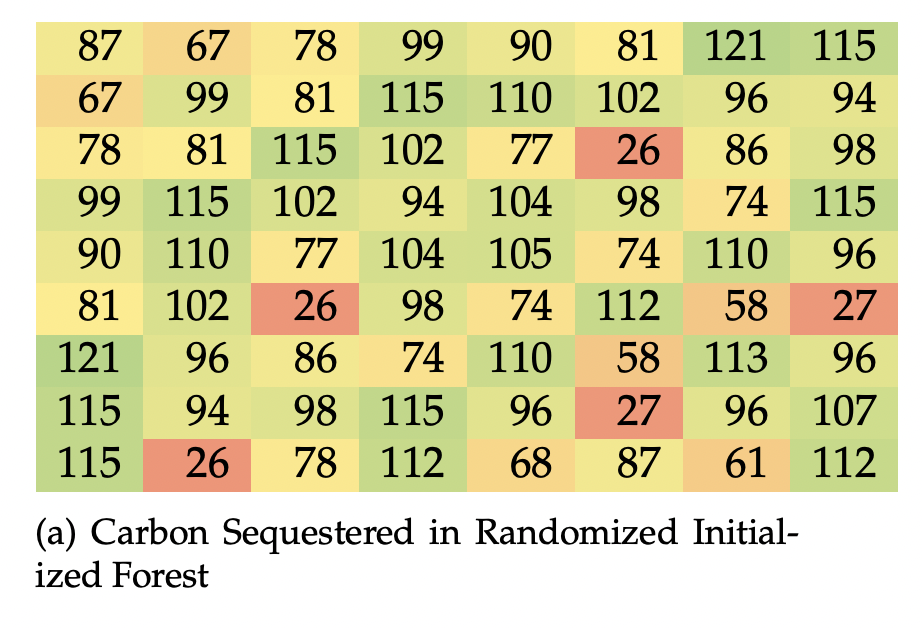}\\
\includegraphics[scale=0.5]{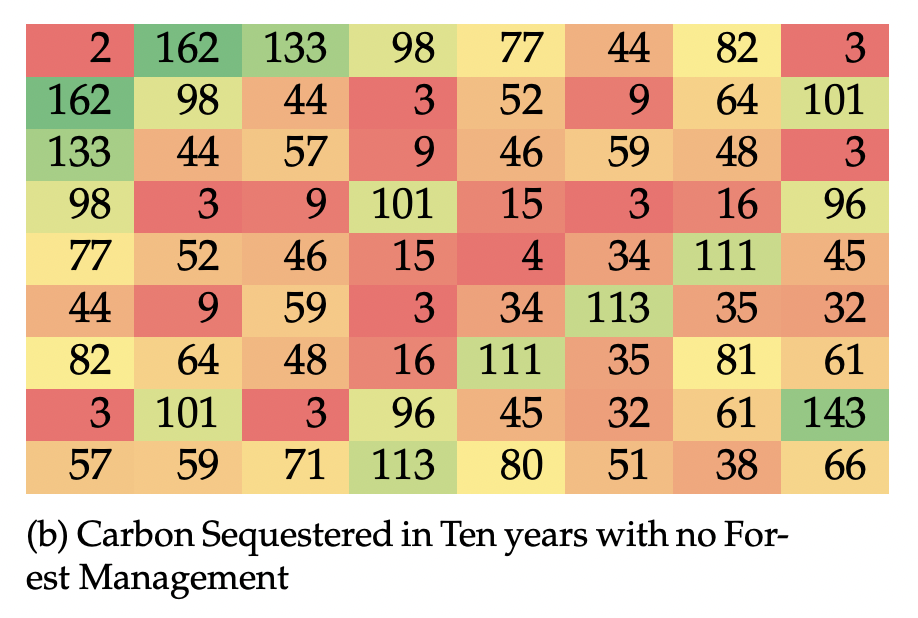}

\pgfplotsset{
compat=newest,
every axis plot/.append style={no marks,thick},
every axis/.style={
  axis lines=middle,
  width=18cm,
  height=5cm,
  }
}

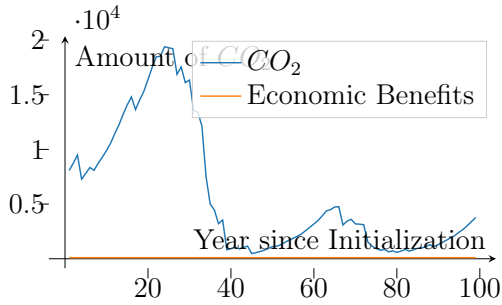
\begin{figure}[H]
  \center
    \begin{tabular}{rrrrrrrrrrrrrrr}
    \begin{tikzpicture}[scale=0.9]

\definecolor{darkgray176}{RGB}{176,176,176}
\definecolor{darkorange25512714}{RGB}{255,127,14}
\definecolor{lightgray204}{RGB}{204,204,204}
\definecolor{steelblue31119180}{RGB}{31,119,180}

\begin{axis}[
width=0.95\linewidth,  
legend cell align={left},
legend style={fill opacity=0.8, draw opacity=1, text opacity=1, draw=lightgray204},
tick align=outside,
tick pos=left,
x grid style={darkgray176},
xlabel={Year since Initialization},
xmin=-3.9, xmax=103.9,
xtick style={color=black},
y grid style={darkgray176},
ylabel={Amount of \(\displaystyle {CO_2}\)},
ymin=-864.625197858425, ymax=20357.1291550269,
ytick style={color=black}
]
\addplot [semithick, steelblue31119180]
table {%
1 8072.24225943224
2 8742.73294564377
3 9484.0572414125
4 7286.77972016987
5 7807.09969398853
6 8347.40779256334
7 8092.20292595111
8 8740.33738682755
9 9293.30562036354
10 9906.60472496984
11 10574.7877444192
12 11448.7121627679
13 12233.8223958949
14 13201.6982162561
15 14075.1598270426
16 14798.7447740596
17 13646.6532452576
18 14530.943065674
19 15303.0306022562
20 16387.2488093513
21 17488.0824317809
22 18462.0255053507
23 18521.6258909435
24 19392.5039571685
25 19298.8124809478
26 19208.1398614639
27 16871.9995495525
28 17538.4402218163
29 16109.4422511748
30 16341.2349162684
31 13578.816890484
32 13412.4260111465
33 12149.6458227123
34 7504.27852050263
35 4999.04083235556
36 4473.85710987846
37 3203.04179824578
38 3545.15445329727
39 825.880069750536
40 934.40858256883
41 1033.50258293993
42 910.252925241917
43 1026.59600493611
44 1161.31582054435
45 476.193684083854
46 557.131166485391
47 659.329830585505
48 776.901182020154
49 934.063730124976
50 1110.13968105615
51 1140.0021391794
52 1272.86444831228
53 1497.08881709338
54 1702.28786941691
55 1877.2488429452
56 2055.85042189487
57 2265.65162111433
58 2569.9111854451
59 2874.73387068327
60 3180.93728235094
61 3495.52949581893
62 3921.26942610989
63 4371.92161834491
64 4487.99056742241
65 4732.62615811783
66 4760.11802129635
67 3087.43141778031
68 3439.26374019205
69 3609.45406687381
70 3185.82266757309
71 3165.32314405603
72 3112.45201791756
73 1532.3093956267
74 1156.25971043496
75 945.085784382379
76 773.622210635139
77 861.074891923924
78 636.540640582275
79 719.505762704063
80 579.670524113276
81 698.911260461125
82 843.255126856193
83 703.682262782287
84 816.95453318943
85 963.237920302843
86 977.812087982421
87 1117.01702115694
88 1289.66787586595
89 1103.83213839635
90 1303.09490048394
91 1503.30976919111
92 1708.0249321155
93 1931.07487756026
94 2134.90529787754
95 2429.93407784558
96 2694.17137771853
97 3039.6016931535
98 3403.30952919728
99 3777.82818167324
};
\addlegendentry{${CO_2}$}
\addplot[semithick, darkorange25512714]
table {%
1 100
2 100
3 100
4 100
5 100
6 100
7 100
8 100
9 100
10 100
11 100
12 100
13 100
14 100
15 100
16 100
17 100
18 100
19 100
20 100
21 100
22 100
23 100
24 100
25 100
26 100
27 100
28 100
29 100
30 100
31 100
32 100
33 100
34 100
35 100
36 100
37 100
38 100
39 100
40 100
41 100
42 100
43 100
44 100
45 100
46 100
47 100
48 100
49 100
50 100
51 100
52 100
53 100
54 100
55 100
56 100
57 100
58 100
59 100
60 100
61 100
62 100
63 100
64 100
65 100
66 100
67 100
68 100
69 100
70 100
71 100
72 100
73 100
74 100
75 100
76 100
77 100
78 100
79 100
80 100
81 100
82 100
83 100
84 100
85 100
86 100
87 100
88 100
89 100
90 100
91 100
92 100
93 100
94 100
95 100
96 100
97 100
98 100
99 100
};
\addlegendentry{Economic Benefits}
\end{axis}
\end{tikzpicture}
    \end{tabular}%
  \label{}
  \caption{Change curve of $CO_2$ and Economic Benefits with time}
\end{figure}%

\subsection{Age-dependent Approach}

Literature has revealed that with increase in time, trees are increasingly susceptible to various environmental cues, including pathogen infection, interspecies competition, and environmental stress. Based on the model, we devise an age based tree logging approach to harvest wood while maintaining forest vitality. Notably, in the Java algorithm, we extend the trees in the aforementioned model to be a $Comparable$ one, and modeled tree growth and carbon sequestered from each year by pushing the resulting ArrayList object onto the Stack. This approach ensured the constancy of comparison for different methods, and further ensured the randomness of initialized status.

We assume that consistently, every year regardless of the environmental status, the tree whose age is close to the imagined death age (1 year from the death expectancy) would be clearly identified and cut completely, thereby providing equal amounts of economic benefits as that of the $CO_2$.

Matplotlib was then utilized to provide a plot of Benefits from both environmental $CO_2$ and wood collection. In comparison with the baseline approach, the scatter plot suggests that due to non-consideration of environmental conditions, continuous wood logging shows temporary decreased overall benefits based on utility function under arid conditions. However, in other times, not only was the forest still a net $CO_2$ sink, but also there had been continuous economic benefits. 

We also confirm the regeneration ability of forests under appropriate logging under the age-selection approach. Color scale follows the same methodology as described in the aforementioned baseline approach. As shown in the page that continues, we see a clear pattern in which a 55-year prediction suggests that tree growth are unencumbered by human logging.

\pgfplotsset{
compat=newest,
every axis plot/.append style={no marks,thick},
every axis/.style={
  axis lines=middle,
  width=8cm,
  height=7cm,
  }
}
\begin{tikzpicture}[scale = 0.9]

\definecolor{crimson2143940}{RGB}{214,39,40}
\definecolor{darkgray176}{RGB}{176,176,176}
\definecolor{darkorange25512714}{RGB}{255,127,14}
\definecolor{forestgreen4416044}{RGB}{44,160,44}
\definecolor{lightgray204}{RGB}{204,204,204}
\definecolor{steelblue31119180}{RGB}{31,119,180}

\begin{axis}[
legend cell align={left},
legend style={at={(0.5,-0.17)},fill opacity=0.1, draw opacity=1, text opacity=1,
draw=lightgray204,anchor=north,legend cell align=left},
tick align=outside,
tick pos=left,
x grid style={darkgray176},
xlabel={Year since Initialization},
xmin=-3.95, xmax=104.95,
xtick style={color=black},
y grid style={darkgray176},
ylabel={Amount of Benefits in All terms},
ymin=-152.200579253359, ymax=3196.21216432053,
ytick style={color=black}
]
\path [draw=steelblue31119180, semithick]
(axis cs:1,0)
--(axis cs:1,0);

\path [draw=steelblue31119180, semithick]
(axis cs:2,1663.13807171692)
--(axis cs:2,1663.13807171692);

\path [draw=steelblue31119180, semithick]
(axis cs:3,0)
--(axis cs:3,0);

\path [draw=steelblue31119180, semithick]
(axis cs:4,0)
--(axis cs:4,0);

\path [draw=steelblue31119180, semithick]
(axis cs:5,983.004928768086)
--(axis cs:5,1331.12134466041);

\path [draw=steelblue31119180, semithick]
(axis cs:6,1153.58964440257)
--(axis cs:6,1654.54553718439);

\path [draw=steelblue31119180, semithick]
(axis cs:7,1039.4252456914)
--(axis cs:7,1751.37449602127);

\path [draw=steelblue31119180, semithick]
(axis cs:8,1227.44068764355)
--(axis cs:8,1227.44068764355);

\path [draw=steelblue31119180, semithick]
(axis cs:9,1444.84988226652)
--(axis cs:9,1444.84988226652);

\path [draw=steelblue31119180, semithick]
(axis cs:10,0)
--(axis cs:10,0);
          
\path [draw=steelblue31119180, semithick]
(axis cs:11,1126.46439098251)
--(axis cs:11,1644.28722508533);

\path [draw=steelblue31119180, semithick]
(axis cs:12,1101.54334893643)
--(axis cs:12,1924.96143388454);

\path [draw=steelblue31119180, semithick]
(axis cs:13,1445.50575631498)
--(axis cs:13,1721.2419005441);

\path [draw=steelblue31119180, semithick]
(axis cs:14,1240.11517233946)
--(axis cs:14,1701.74158833595);

\path [draw=steelblue31119180, semithick]
(axis cs:15,1661.02420130296)
--(axis cs:15,2266.95790061589);

\path [draw=steelblue31119180, semithick]
(axis cs:16,0)
--(axis cs:16,0);

\path [draw=steelblue31119180, semithick]
(axis cs:17,0)
--(axis cs:17,0);

\path [draw=steelblue31119180, semithick]
(axis cs:18,0)
--(axis cs:18,0);

\path [draw=steelblue31119180, semithick]
(axis cs:19,0)
--(axis cs:19,0);

\path [draw=steelblue31119180, semithick]
(axis cs:20,0)
--(axis cs:20,0);

\path [draw=steelblue31119180, semithick]
(axis cs:21,0)
--(axis cs:21,0);

\path [draw=steelblue31119180, semithick]
(axis cs:22,0)
--(axis cs:22,0);

\path [draw=steelblue31119180, semithick]
(axis cs:23,0)
--(axis cs:23,0);

\path [draw=steelblue31119180, semithick]
(axis cs:24,0)
--(axis cs:24,0);

\path [draw=steelblue31119180, semithick]
(axis cs:25,1303.8088785215)
--(axis cs:25,1303.8088785215);

\path [draw=steelblue31119180, semithick]
(axis cs:26,0)
--(axis cs:26,0);

\path [draw=steelblue31119180, semithick]
(axis cs:27,0)
--(axis cs:27,0);

\path [draw=steelblue31119180, semithick]
(axis cs:28,0)
--(axis cs:28,0);

\path [draw=steelblue31119180, semithick]
(axis cs:29,0)
--(axis cs:29,0);

\path [draw=steelblue31119180, semithick]
(axis cs:30,1636.65851013301)
--(axis cs:30,2111.91197704178);

\path [draw=steelblue31119180, semithick]
(axis cs:31,0)
--(axis cs:31,0);

\path [draw=steelblue31119180, semithick]
(axis cs:32,0)
--(axis cs:32,0);

\path [draw=steelblue31119180, semithick]
(axis cs:33,1636.62957698568)
--(axis cs:33,1713.66649008047);

\path [draw=steelblue31119180, semithick]
(axis cs:34,0)
--(axis cs:34,0);

\path [draw=steelblue31119180, semithick]
(axis cs:35,0)
--(axis cs:35,0);

\path [draw=steelblue31119180, semithick]
(axis cs:36,0)
--(axis cs:36,0);

\path [draw=steelblue31119180, semithick]
(axis cs:37,0)
--(axis cs:37,0);

\path [draw=steelblue31119180, semithick]
(axis cs:38,0)
--(axis cs:38,0);

\path [draw=steelblue31119180, semithick]
(axis cs:39,0)
--(axis cs:39,0);

\path [draw=steelblue31119180, semithick]
(axis cs:40,1471.34952413386)
--(axis cs:40,1471.34952413386);

\path [draw=steelblue31119180, semithick]
(axis cs:41,0)
--(axis cs:41,0);

\path [draw=steelblue31119180, semithick]
(axis cs:42,1679.33933166285)
--(axis cs:42,1861.10014656978);

\path [draw=steelblue31119180, semithick]
(axis cs:43,1178.98884470192)
--(axis cs:43,2176.39952849785);

\path [draw=steelblue31119180, semithick]
(axis cs:44,1001.34181667189)
--(axis cs:44,1580.38653451872);

\path [draw=steelblue31119180, semithick]
(axis cs:45,2123.51292285315)
--(axis cs:45,2123.51292285315);

\path [draw=steelblue31119180, semithick]
(axis cs:46,1699.44191616018)
--(axis cs:46,2108.54434348441);

\path [draw=steelblue31119180, semithick]
(axis cs:47,2056.84699227761)
--(axis cs:47,2056.84699227761);

\path [draw=steelblue31119180, semithick]
(axis cs:48,0)
--(axis cs:48,0);

\path [draw=steelblue31119180, semithick]
(axis cs:49,1318.57025257634)
--(axis cs:49,1318.57025257634);

\path [draw=steelblue31119180, semithick]
(axis cs:50,1885.34742844976)
--(axis cs:50,1885.34742844976);

\path [draw=steelblue31119180, semithick]
(axis cs:51,1142.62830943027)
--(axis cs:51,1457.24475721389);

\path [draw=steelblue31119180, semithick]
(axis cs:52,1024.72095048308)
--(axis cs:52,2071.17756224085);

\path [draw=steelblue31119180, semithick]
(axis cs:53,1504.70600187078)
--(axis cs:53,1504.70600187078);

\path [draw=steelblue31119180, semithick]
(axis cs:54,0)
--(axis cs:54,0);

\path [draw=steelblue31119180, semithick]
(axis cs:55,1691.35310161742)
--(axis cs:55,2854.00440827118);

\path [draw=steelblue31119180, semithick]
(axis cs:56,892.516632407217)
--(axis cs:56,892.516632407217);

\path [draw=steelblue31119180, semithick]
(axis cs:57,0)
--(axis cs:57,0);

\path [draw=steelblue31119180, semithick]
(axis cs:58,0)
--(axis cs:58,0);

\path [draw=steelblue31119180, semithick]
(axis cs:59,0)
--(axis cs:59,0);

\path [draw=steelblue31119180, semithick]
(axis cs:60,0)
--(axis cs:60,0);

\path [draw=steelblue31119180, semithick]
(axis cs:61,1472.23853811411)
--(axis cs:61,1472.23853811411);

\path [draw=steelblue31119180, semithick]
(axis cs:62,0)
--(axis cs:62,0);

\path [draw=steelblue31119180, semithick]
(axis cs:63,1025.78066615985)
--(axis cs:63,1662.68750741142);

\path [draw=steelblue31119180, semithick]
(axis cs:64,0)
--(axis cs:64,0);

\path [draw=steelblue31119180, semithick]
(axis cs:65,0)
--(axis cs:65,0);

\path [draw=steelblue31119180, semithick]
(axis cs:66,0)
--(axis cs:66,0);

\path [draw=steelblue31119180, semithick]
(axis cs:67,0)
--(axis cs:67,0);

\path [draw=steelblue31119180, semithick]
(axis cs:68,0)
--(axis cs:68,0);

\path [draw=steelblue31119180, semithick]
(axis cs:69,0)
--(axis cs:69,0);

\path [draw=steelblue31119180, semithick]
(axis cs:70,0)
--(axis cs:70,0);

\path [draw=steelblue31119180, semithick]
(axis cs:71,974.407403733886)
--(axis cs:71,1016.71592799825);

\path [draw=steelblue31119180, semithick]
(axis cs:72,0)
--(axis cs:72,0);

\path [draw=steelblue31119180, semithick]
(axis cs:73,0)
--(axis cs:73,0);

\path [draw=steelblue31119180, semithick]
(axis cs:74,0)
--(axis cs:74,0);

\path [draw=steelblue31119180, semithick]
(axis cs:75,0)
--(axis cs:75,0);

\path [draw=steelblue31119180, semithick]
(axis cs:76,0)
--(axis cs:76,0);

\path [draw=steelblue31119180, semithick]
(axis cs:77,0)
--(axis cs:77,0);

\path [draw=steelblue31119180, semithick]
(axis cs:78,0)
--(axis cs:78,0);

\path [draw=steelblue31119180, semithick]
(axis cs:79,0)
--(axis cs:79,0);

\path [draw=steelblue31119180, semithick]
(axis cs:80,1724.62487036431)
--(axis cs:80,1724.62487036431);

\path [draw=steelblue31119180, semithick]
(axis cs:81,0)
--(axis cs:81,0);

\path [draw=steelblue31119180, semithick]
(axis cs:82,1673.36071060442)
--(axis cs:82,1673.36071060442);

\path [draw=steelblue31119180, semithick]
(axis cs:83,998.367472255892)
--(axis cs:83,1423.8208011141);

\path [draw=steelblue31119180, semithick]
(axis cs:84,1325.57872380704)
--(axis cs:84,1606.23106346958);

\path [draw=steelblue31119180, semithick]
(axis cs:85,0)
--(axis cs:85,0);

\path [draw=steelblue31119180, semithick]
(axis cs:86,0)
--(axis cs:86,0);

\path [draw=steelblue31119180, semithick]
(axis cs:87,1819.5090894603)
--(axis cs:87,1819.5090894603);

\path [draw=steelblue31119180, semithick]
(axis cs:88,1114.01887216697)
--(axis cs:88,1114.01887216697);

\path [draw=steelblue31119180, semithick]
(axis cs:89,1336.8991699006)
--(axis cs:89,1765.65520556479);

\path [draw=steelblue31119180, semithick]
(axis cs:90,1386.90271990905)
--(axis cs:90,1490.0867791538);

\path [draw=steelblue31119180, semithick]
(axis cs:91,969.636657104435)
--(axis cs:91,1914.11126155094);

\path [draw=steelblue31119180, semithick]
(axis cs:92,955.582622707674)
--(axis cs:92,1568.31288574753);

\path [draw=steelblue31119180, semithick]
(axis cs:93,1023.57923031527)
--(axis cs:93,1062.76680796887);

\path [draw=steelblue31119180, semithick]
(axis cs:94,0)
--(axis cs:94,0);

\path [draw=steelblue31119180, semithick]
(axis cs:95,1401.5041606228)
--(axis cs:95,1401.5041606228);

\path [draw=steelblue31119180, semithick]
(axis cs:96,0)
--(axis cs:96,0);

\path [draw=steelblue31119180, semithick]
(axis cs:97,0)
--(axis cs:97,0);

\path [draw=steelblue31119180, semithick]
(axis cs:98,0)
--(axis cs:98,0);

\path [draw=steelblue31119180, semithick]
(axis cs:99,0)
--(axis cs:99,0);

\path [draw=steelblue31119180, semithick]
(axis cs:100,0)
--(axis cs:100,0);

\addplot [semithick, darkorange25512714, dotted]
table {%
1 244.773646248868
2 273.01940896638
3 290.115565733874
4 245.747273066934
5 278.431555227114
6 254.168635456588
7 236.384321950192
8 268.496043333025
9 283.097266097631
10 261.826024665569
11 259.891266885025
12 152.871214503079
13 134.971474005478
14 152.425775134169
15 143.946897842313
16 161.243757982245
17 184.633743332064
18 180.901017696207
19 198.913421746619
20 170.131236711121
21 191.91051340003
22 212.244584341833
23 147.324966894854
24 169.176117259374
25 197.577818710356
26 169.316686142235
27 190.285672244231
28 211.293613568349
29 157.37385529443
30 178.286950034967
31 204.345046888717
32 228.068879798274
33 250.245077165305
34 278.735320453857
35 220.231442778738
36 251.132752657749
37 250.34470349435
38 227.17832042181
39 249.837745753102
40 221.962502141324
41 224.798032979351
42 256.383212613326
43 270.117382024028
44 253.509377317571
45 239.983744168933
46 161.967333958905
47 136.189004810724
48 156.255770004696
49 160.38810223602
50 177.542778217287
51 199.753147010211
52 187.869140793097
53 217.462652978229
54 167.508267782768
55 190.007176795995
56 211.944091268009
57 157.813980499126
58 177.605591400632
59 203.862215582548
60 180.135381331664
61 205.391019717966
62 226.21612175368
63 159.788628525962
64 180.195996375264
65 204.546193143718
66 230.80440768081
67 261.361645742896
68 296.46581387738
69 207.762498025377
70 234.716071491939
71 239.562913892991
72 241.829866331935
73 273.764317428783
74 261.007799634763
75 252.46160296123
76 281.743156665264
77 307.96652814755
78 310.049661729352
79 291.20371672972
80 178.18282573999
81 163.507454602781
82 181.104034980109
83 155.809513010994
84 173.761618275203
85 197.646480049537
86 186.011050883434
87 214.376810365885
88 149.840341386799
89 165.847454729411
90 186.625060603443
91 146.714479262347
92 166.573408535487
93 189.906539144146
94 174.779538368888
95 198.194881517943
96 214.297256512407
97 161.596591316844
98 184.38004378655
99 206.627032201509
100 0
};
\addlegendentry{${CO_2}$}
\addplot [semithick, forestgreen4416044]
table {%
1 807.224225943224
2 874.273294564377
3 948.40572414125
4 728.677972016987
5 780.709969398853
6 834.740779256334
7 809.220292595111
8 874.033738682755
9 929.330562036354
10 990.660472496984
11 1057.47877444192
12 1144.87121627679
13 1223.38223958949
14 1320.16982162561
15 1407.51598270426
16 1479.87447740596
17 1364.66532452576
18 1453.0943065674
19 1530.30306022562
20 1638.72488093513
21 1748.80824317809
22 1846.20255053507
23 1852.16258909435
24 1939.25039571685
25 1929.88124809478
26 1920.81398614639
27 1687.19995495525
28 1753.84402218163
29 1610.94422511748
30 1634.12349162684
31 1357.8816890484
32 1341.24260111465
33 1214.96458227123
34 750.427852050263
35 499.904083235556
36 447.385710987846
37 320.304179824578
38 354.515445329727
39 82.5880069750536
40 93.440858256883
41 103.350258293993
42 91.0252925241917
43 102.659600493611
44 116.131582054435
45 47.6193684083854
46 55.7131166485391
47 65.9329830585505
48 77.6901182020154
49 93.4063730124976
50 111.013968105615
51 114.00021391794
52 127.286444831228
53 149.708881709338
54 170.228786941691
55 187.72488429452
56 205.585042189487
57 226.565162111433
58 256.99111854451
59 287.473387068327
60 318.093728235094
61 349.552949581893
62 392.126942610989
63 437.192161834491
64 448.799056742241
65 473.262615811783
66 476.011802129635
67 308.743141778031
68 343.926374019205
69 360.945406687381
70 318.582266757309
71 316.532314405603
72 311.245201791756
73 153.23093956267
74 115.625971043496
75 94.5085784382379
76 77.3622210635139
77 86.1074891923924
78 63.6540640582275
79 71.9505762704063
80 57.9670524113276
81 69.8911260461125
82 84.3255126856193
83 70.3682262782287
84 81.695453318943
85 96.3237920302843
86 97.7812087982421
87 111.701702115694
88 128.966787586595
89 110.383213839635
90 130.309490048394
91 150.330976919111
92 170.80249321155
93 193.107487756026
94 213.490529787754
95 242.993407784558
96 269.417137771853
97 303.96016931535
98 340.330952919728
99 377.782818167324
100 377.782818167324
};
\addlegendentry{Benefits estimate of no action}
\addplot [semithick, crimson2143940, dashed]
table {%
1 244.773646248868
2 1936.1574806833
3 290.115565733874
4 245.747273066934
5 1609.55289988753
6 1908.71417264098
7 1987.75881797146
8 1495.93673097657
9 1727.94714836415
10 261.826024665569
11 1904.17849197036
12 2077.83264838762
13 1856.21337454958
14 1854.16736347012
15 2410.9047984582
16 161.243757982245
17 184.633743332064
18 180.901017696207
19 198.913421746619
20 170.131236711121
21 191.91051340003
22 212.244584341833
23 147.324966894854
24 169.176117259374
25 1501.38669723186
26 169.316686142235
27 190.285672244231
28 211.293613568349
29 157.37385529443
30 2290.19892707674
31 204.345046888717
32 228.068879798274
33 1963.91156724577
34 278.735320453857
35 220.231442778738
36 251.132752657749
37 250.34470349435
38 227.17832042181
39 249.837745753102
40 1693.31202627518
41 224.798032979351
42 2117.48335918311
43 2446.51691052188
44 1833.89591183629
45 2363.49666702208
46 2270.51167744332
47 2193.03599708834
48 156.255770004696
49 1478.95835481236
50 2062.89020666705
51 1656.9979042241
52 2259.04670303395
53 1722.16865484901
54 167.508267782768
55 3044.01158506717
56 1104.46072367523
57 157.813980499126
58 177.605591400632
59 203.862215582548
60 180.135381331664
61 1677.62955783208
62 226.21612175368
63 1822.47613593738
64 180.195996375264
65 204.546193143718
66 230.80440768081
67 261.361645742896
68 296.46581387738
69 207.762498025377
70 234.716071491939
71 1256.27884189124
72 241.829866331935
73 273.764317428783
74 261.007799634763
75 252.46160296123
76 281.743156665264
77 307.96652814755
78 310.049661729352
79 291.20371672972
80 1902.8076961043
81 163.507454602781
82 1854.46474558453
83 1579.6303141251
84 1779.99268174478
85 197.646480049537
86 186.011050883434
87 2033.88589982618
88 1263.85921355376
89 1931.5026602942
90 1676.71183975725
91 2060.82574081329
92 1734.88629428301
93 1252.67334711301
94 174.779538368888
95 1599.69904214074
96 214.297256512407
97 161.596591316844
98 184.38004378655
99 206.627032201509
100 0
};
\addlegendentry{dotted estimate of Age-specified Tree Loggin}
\addplot [semithick, steelblue31119180, mark=triangle*, mark size=3, mark options={solid}, only marks]
table {%
1 0
2 1663.13807171692
3 0
4 0
5 1157.06313671425
6 1404.06759079348
7 1395.39987085634
8 1227.44068764355
9 1444.84988226652
10 0
11 1385.37580803392
12 1513.25239141048
13 1583.37382842954
14 1470.92838033771
15 1963.99105095942
16 0
17 0
18 0
19 0
20 0
21 0
22 0
23 0
24 0
25 1303.8088785215
26 0
27 0
28 0
29 0
30 1874.28524358739
31 0
32 0
33 1675.14803353307
34 0
35 0
36 0
37 0
38 0
39 0
40 1471.34952413386
41 0
42 1770.21973911632
43 1677.69418659989
44 1290.86417559531
45 2123.51292285315
46 1903.99312982229
47 2056.84699227761
48 0
49 1318.57025257634
50 1885.34742844976
51 1299.93653332208
52 1547.94925636197
53 1504.70600187078
54 0
55 2272.6787549443
56 892.516632407217
57 0
58 0
59 0
60 0
61 1472.23853811411
62 0
63 1344.23408678563
64 0
65 0
66 0
67 0
68 0
69 0
70 0
71 995.561665866069
72 0
73 0
74 0
75 0
76 0
77 0
78 0
79 0
80 1724.62487036431
81 0
82 1673.36071060442
83 1211.094136685
84 1465.90489363831
85 0
86 0
87 1819.5090894603
88 1114.01887216697
89 1551.2771877327
90 1438.49474953143
91 1441.87395932769
92 1261.9477542276
93 1043.17301914207
94 0
95 1401.5041606228
96 0
97 0
98 0
99 0
100 0
};
\addlegendentry{Economic Benefits with Standard Deviation}
\end{axis}
\end{tikzpicture}

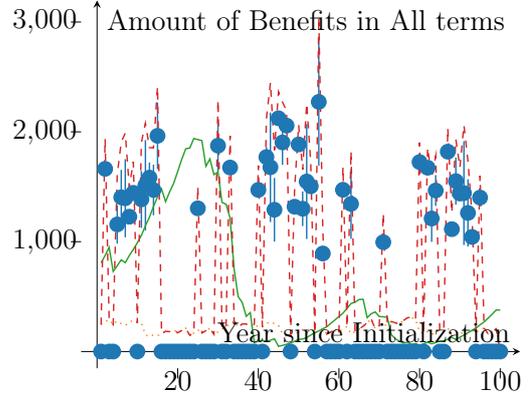
\captionof{figure}{Economic Benefits and Carbon Sequestered versus Baseline. In this figure, dots represent economic benefits, dashed red lines benefits obtained from age-specified tree logging activities, and solid line represents net carbon benefits with no cation.}


\subsection{Indications and Contraindications for Forest Management Plan - Sensitivity and Robustness Analysis}

We adjust the following variables used in our algorithm, and obtain the following data for understanding of proper use of timber cutting as management strategies. Four different forest were evaluated based on existing literature on environmental science, and expected benefits as compared to baseline approach was summarized.
\subsubsection{Effects of Harvest Age concerning Humidity and Harvest Percentage}

We modeled three different hypothetical situations that relates to harvest age:
(i) We want to understand whether this logging approach would be effective in different drought possibility, with step wise progression of 0.3. This has its contemporary implications in that with the advent of global warming, extreme weathers have become more likely and would therefore impact the robustness of prior climate data's use on timber harvesting strategies.

\begin{figure}[h]
  \center
    \begin{tabular}{rrrrrrrrrrrrrrr}
\begin{tikzpicture}[scale=0.4]

\definecolor{darkgray176}{RGB}{176,176,176}
\definecolor{goldenrod1911910}{RGB}{191,191,0}
\definecolor{green01270}{RGB}{0,127,0}
\definecolor{lightgray204}{RGB}{204,204,204}

\begin{axis}[
legend cell align={left},
legend style={at={(0.5,-0.17)},fill opacity=0.1, draw opacity=1, text opacity=1,
draw=lightgray204,anchor=north,legend cell align=left},
tick align=outside,
tick pos=left,
x grid style={darkgray176},
xlabel={Distance of Wood Harvest Cutoff Age from Maximum Age},
xmin=-1.45, xmax=52.45,
xtick style={color=black},
y grid style={darkgray176},
ylabel={Amount of Benefits},
ymin=13654208121.5386, ymax=464157425615.392,
ytick style={color=black}
]
\addplot [semithick, blue]
table {%
1 443680006638.399
2 443622901233.17
3 413317799182.743
4 400421245081.667
5 383697391361.598
6 359506855598.974
7 361739185883.744
8 355355165420.473
9 356113182348.264
10 391888587766.148
11 386701882906.777
12 367790595410.762
13 282224096432.954
14 272416758758.057
15 269029835537.739
16 292988056650.793
17 301070147319.785
18 223069501079.725
19 219649074064.487
20 265872039446.587
21 269994948937.517
22 179306782378.55
23 185346606057.195
24 241839770957.162
25 171081251747.042
26 189117437970.704
27 206452096914.933
28 151179846259.296
29 186154060385.361
30 141646188575.231
31 163428921961.907
32 127923052874.116
33 148951061816.929
34 131780219892.036
35 131149481863.004
36 102258359880.036
37 133644924156.054
38 92820780991.5555
39 101683028906.363
40 115884762971.941
41 85560743486.4359
42 90999234122.4494
43 83739082770.4887
44 84265684954.4832
45 74711416943.8733
46 67234127242.9231
47 65744740842.0575
48 58543637252.8276
49 65229053945.0929
50 59128860940.8995
};
\addlegendentry{0.4 Harvest, wet}
\addplot [semithick, green01270]
table {%
1 245101679625.722
2 244609935813.843
3 251568277301.712
4 236711728702.8
5 223265829547.107
6 233549978537.232
7 207580935337.673
8 203201825635.392
9 207240410819.949
10 208217750730.819
11 234543495588.436
12 206836687657.287
13 156772696483.478
14 153109873467.192
15 157309774823.549
16 168801550086.054
17 184342984298.707
18 151313082121.415
19 126087845155.725
20 140234893512.8
21 154894835010.391
22 119968930808.181
23 107057504126.327
24 141142518066.374
25 114319197120.938
26 107071302356.054
27 123910104320.802
28 86718030543.88
29 104508584112.777
30 88475762942.6818
31 92488427805.6131
32 90860183872.8947
33 79490357394.1856
34 82275287112.5708
35 75852234934.0887
36 68825764851.3832
37 70834964868.4534
38 51409002706.769
39 61697492215.7924
40 63213234488.6624
41 47659846584.2714
42 53687666408.3327
43 50391724681.8231
44 52323712285.6789
45 44131706042.3563
46 37387295589.5217
47 38409638783.64
48 37380758438.9806
49 38321478270.7807
50 35172174662.3122
};
\addlegendentry{0.4 Harvest, wildfire threshold = 0.3}
\addplot [semithick, goldenrod1911910]
table {%
1 242275781271.206
2 235824238344.243
3 227741303786.987
4 218687640079.961
5 237773315155.843
6 221331910941.171
7 210556539833.582
8 202261039555.602
9 204017512621.444
10 216272931680.172
11 219125958519.803
12 219345377709.679
13 170124049024.071
14 152542635786.358
15 155498958316.042
16 172272488912.505
17 203840878190.252
18 149736463390.799
19 120550834505.036
20 155239771903.507
21 175259721801.824
22 111428375451.435
23 107365852803.845
24 137955375781.548
25 113152719005.056
26 98390381161.6064
27 138560524177.837
28 82853924758.4016
29 114163713586.235
30 94637729968.8647
31 78763252769.0922
32 84629523871.9252
33 74832926021.8475
34 89670302859.6702
35 74492914163.4632
36 68238361605.1225
37 75260978334.3486
38 50454101689.8433
39 65485353594.854
40 59912416406.9254
41 47892440672.0866
42 50071255126.2164
43 52983354189.9907
44 45556672863.5467
45 45221279063.2047
46 43943532672.4371
47 37067832038.0269
48 38135226617.5724
49 35290579455.2101
50 34131627098.5319
};
\addlegendentry{0.4 Harvest, wildfire threshold = 0.6}
\addplot [semithick, red]
table {%
1 232929647218.246
2 256488977139.205
3 233373476883.286
4 234754692227.757
5 232468433238.284
6 226892535168.927
7 231620293299.508
8 209493683334.625
9 199346262683.888
10 211219644744.121
11 231241645647.385
12 208024818726.884
13 163456407214.283
14 160240603773.226
15 158324309797.262
16 180969279475.721
17 197232983336.269
18 131290525882.158
19 136098253049.09
20 141601975481.768
21 172980462412.861
22 117515925227.2
23 119099224203.282
24 146252278547.591
25 94134054189.8066
26 102907537369.875
27 134110330566.203
28 83575818380.0307
29 121685579989.208
30 91280223317.9831
31 90607699656.6335
32 84885115260.271
33 75226177109.717
34 78217972954.6845
35 81225209514.0605
36 61681181621.7421
37 78832287978.0501
38 53727586126.653
39 59572173257.3998
40 61620987420.7423
41 51149767714.7332
42 44114249909.242
43 54175053476.4558
44 55968136491.1311
45 48367485568.7374
46 40344027753.0842
47 41469082523.9559
48 34647205021.3687
49 34526333841.328
50 36089507999.469
};
\addlegendentry{0.4 Harvest, wildfire threshold = 0.9}
\addplot [semithick, blue]
table {%
1 350432687032.333
2 350432687032.333
3 350432687032.333
4 350432687032.333
5 350432687032.333
6 350432687032.333
7 350432687032.333
8 350432687032.333
9 350432687032.333
10 350432687032.333
11 350432687032.333
12 350432687032.333
13 350432687032.333
14 350432687032.333
15 350432687032.333
16 350432687032.333
17 350432687032.333
18 350432687032.333
19 350432687032.333
20 350432687032.333
21 350432687032.333
22 350432687032.333
23 350432687032.333
24 350432687032.333
25 350432687032.333
26 350432687032.333
27 350432687032.333
28 350432687032.333
29 350432687032.333
30 350432687032.333
31 350432687032.333
32 350432687032.333
33 350432687032.333
34 350432687032.333
35 350432687032.333
36 350432687032.333
37 350432687032.333
38 350432687032.333
39 350432687032.333
40 350432687032.333
41 350432687032.333
42 350432687032.333
43 350432687032.333
44 350432687032.333
45 350432687032.333
46 350432687032.333
47 350432687032.333
48 350432687032.333
49 350432687032.333
50 350432687032.333
};
\addlegendentry{Baseline, wet}
\addplot [semithick, green01270]
table {%
1 178173449755.278
2 178173449755.278
3 178173449755.278
4 178173449755.278
5 178173449755.278
6 178173449755.278
7 178173449755.278
8 178173449755.278
9 178173449755.278
10 178173449755.278
11 178173449755.278
12 178173449755.278
13 178173449755.278
14 178173449755.278
15 178173449755.278
16 178173449755.278
17 178173449755.278
18 178173449755.278
19 178173449755.278
20 178173449755.278
21 178173449755.278
22 178173449755.278
23 178173449755.278
24 178173449755.278
25 178173449755.278
26 178173449755.278
27 178173449755.278
28 178173449755.278
29 178173449755.278
30 178173449755.278
31 178173449755.278
32 178173449755.278
33 178173449755.278
34 178173449755.278
35 178173449755.278
36 178173449755.278
37 178173449755.278
38 178173449755.278
39 178173449755.278
40 178173449755.278
41 178173449755.278
42 178173449755.278
43 178173449755.278
44 178173449755.278
45 178173449755.278
46 178173449755.278
47 178173449755.278
48 178173449755.278
49 178173449755.278
50 178173449755.278
};
\addlegendentry{Baseline, wildfire threshold = 0.3}
\addplot [semithick, goldenrod1911910]
table {%
1 175783312200.995
2 175783312200.995
3 175783312200.995
4 175783312200.995
5 175783312200.995
6 175783312200.995
7 175783312200.995
8 175783312200.995
9 175783312200.995
10 175783312200.995
11 175783312200.995
12 175783312200.995
13 175783312200.995
14 175783312200.995
15 175783312200.995
16 175783312200.995
17 175783312200.995
18 175783312200.995
19 175783312200.995
20 175783312200.995
21 175783312200.995
22 175783312200.995
23 175783312200.995
24 175783312200.995
25 175783312200.995
26 175783312200.995
27 175783312200.995
28 175783312200.995
29 175783312200.995
30 175783312200.995
31 175783312200.995
32 175783312200.995
33 175783312200.995
34 175783312200.995
35 175783312200.995
36 175783312200.995
37 175783312200.995
38 175783312200.995
39 175783312200.995
40 175783312200.995
41 175783312200.995
42 175783312200.995
43 175783312200.995
44 175783312200.995
45 175783312200.995
46 175783312200.995
47 175783312200.995
48 175783312200.995
49 175783312200.995
50 175783312200.995
};
\addlegendentry{Baseline, wildfire threshold = 0.6}
\addplot [semithick, red]
table {%
1 190742682836.339
2 190742682836.339
3 190742682836.339
4 190742682836.339
5 190742682836.339
6 190742682836.339
7 190742682836.339
8 190742682836.339
9 190742682836.339
10 190742682836.339
11 190742682836.339
12 190742682836.339
13 190742682836.339
14 190742682836.339
15 190742682836.339
16 190742682836.339
17 190742682836.339
18 190742682836.339
19 190742682836.339
20 190742682836.339
21 190742682836.339
22 190742682836.339
23 190742682836.339
24 190742682836.339
25 190742682836.339
26 190742682836.339
27 190742682836.339
28 190742682836.339
29 190742682836.339
30 190742682836.339
31 190742682836.339
32 190742682836.339
33 190742682836.339
34 190742682836.339
35 190742682836.339
36 190742682836.339
37 190742682836.339
38 190742682836.339
39 190742682836.339
40 190742682836.339
41 190742682836.339
42 190742682836.339
43 190742682836.339
44 190742682836.339
45 190742682836.339
46 190742682836.339
47 190742682836.339
48 190742682836.339
49 190742682836.339
50 190742682836.339
};
\addlegendentry{Baseline, wildfire threshold = 0.9}
\end{axis}

\end{tikzpicture}

    \end{tabular}%
  \label{tab:addlabel}%
  \caption{Change curve of benefit and cutoff age with various wetness and wildfire threshold}
\end{figure}%

The above trend lines yield several important conclusions which visualizes the following mathematical trends:

\begin{enumerate}
\item[(A)] The higher the cutoff age for harvesting, the more likely it is going to yield more benefits than no clear cutting of the trees.

\item[(B)] The humidity does not necessarily guarantee a lower cutoff age to be beneficial. We see that at wetness equivalent as 0.6, meaning that when environmental humidity is 0.6 out of a total of 1, the trees in a $3\times3$ grid would go on fire, which is when the threshold of maximum age types is actually the largest. 

\item[(C)] The model effectively pictures a similar trend line, albeit providing the critical information as to the non-linear dependency of age groups versus the aridity of the environment.
\end{enumerate}
We also are interested in investigating variable factors of harvest percentage. Given the algorithm, our approach is to determine a random number and compare it to \(1-\text{harvest percentage}\). Data was run on matplotlib and obtained as below. This test primarily serves as a sensitivity approach towards the values, and we recognize a decrease in the intersection between age of trees before its death and the percentage we count. This complements and corroborates the methodology depicted in part (i) of this section.

\pgfplotsset{
compat=newest,
every axis plot/.append style={no marks,thick},
every axis/.style={
  axis lines=middle,
  width=14cm,
  height=6cm,
  }
}
\begin{figure}[h]
  \center
    \begin{tabular}{rrrrrrrrrrrrrrr}
\begin{tikzpicture}[scale = 0.4]

\definecolor{darkgray176}{RGB}{176,176,176}
\definecolor{goldenrod1911910}{RGB}{191,191,0}
\definecolor{green01270}{RGB}{0,127,0}
\definecolor{lightgray204}{RGB}{204,204,204}

\begin{axis}[
legend cell align={left},
legend style={at={(1.0,0.47)},fill opacity=0.1, draw opacity=1, text opacity=1,draw=lightgray204,anchor=west,legend cell align=left},
tick align=outside,
tick pos=left,
x grid style={darkgray176},
xlabel={Wood Harvest Cutoff Age},
xmin=-1.45, xmax=52.45,
xtick style={color=black},
y grid style={darkgray176},
ylabel={Amount of Benefits},
ymin=17347404490.8928, ymax=275354872370.814,
ytick style={color=black}
]
\addplot [semithick, blue]
table {%
1 246727244155.571
2 236290928466.779
3 245654028259.385
4 236090184978.896
5 245571501840.629
6 235999740615.102
7 234779539324.036
8 231437059750.474
9 211246450089.318
10 220894771748.551
11 226923985595.761
12 232119946191.429
13 235451556895.96
14 224450883069.604
15 238256482256.857
16 230026588561.98
17 236319619574.325
18 195367224897.527
19 190747999303.09
20 165133433865.715
21 175040834732.846
22 191139868271.513
23 214954547609.129
24 193690095439.658
25 204766929240.917
26 168153679188.521
27 152521041707.978
28 177202689016.277
29 148658664009.346
30 132865890155.13
31 129005652019.382
32 148571843129.324
33 141903237627.044
34 117096928502.411
35 124668551210.869
36 122635401584.028
37 125232808229.807
38 116346262709.397
39 109948370178.472
40 121033292620.99
41 100502687667.22
42 87378669981.9104
43 110804649481.478
44 132891376322.457
45 90491507149.8184
46 95866574429.8211
47 120088891724.415
48 60427021756.0841
49 86449250730.7228
50 82593914199.3261
};
\addlegendentry{0.1 Harvest}
\addplot [semithick, green01270]
table {%
1 246161887770.705
2 244601177998.184
3 231682478886.574
4 257196922887.836
5 223720193620.897
6 214768228843.352
7 214824748011.283
8 215257454378.227
9 209439639146.494
10 210919867297.376
11 219881677442.482
12 228816578754.657
13 198076990392.911
14 173580291701.197
15 158530664372.798
16 164794984008.707
17 194681823797.618
18 171769916029.662
19 121292394932.86
20 133207645284.707
21 155288629200.553
22 151062941669.399
23 126938058730.346
24 121467670373.5
25 139227619644.851
26 104369330306.718
27 106236633118.017
28 113298552410.573
29 85924064649.3447
30 108989530025.621
31 82404868398.8925
32 89903669874.698
33 81598365650.7961
34 84449916658.1702
35 78723647997.9967
36 75624885275.74
37 66024954139.5072
38 73481057911.6861
39 61175917714.876
40 57121063099.2337
41 63398866302.5817
42 54046307913.0241
43 51219514989.5526
44 50392968816.0044
45 47223362876.1446
46 46279803354.68
47 39810533521.0956
48 39505709163.4815
49 34811631788.8554
50 32715149773.1584
};
\addlegendentry{0.3 Harvest}
\addplot [semithick, goldenrod1911910]
table {%
1 263627260194.454
2 238692316483.695
3 240026839829.583
4 234283473642.828
5 233534304030.227
6 210454749340.711
7 202489059019.744
8 212982524586.599
9 217814890193.05
10 222866668612.125
11 221511290108.199
12 177270550306.99
13 163429813361.799
14 172820607008.229
15 163961609898.154
16 181559214192.828
17 152232534258.805
18 119248692526.357
19 128177953115.782
20 166807578889.723
21 116527782724.257
22 105906970614.276
23 144525089416.334
24 121615464259.949
25 91793364872.9228
26 133140025988.37
27 88172421851.3126
28 96756085873.4035
29 96743023909.0643
30 79997708543.8494
31 100074674969.361
32 77044405719.7727
33 90443985661.4653
34 65620072461.1949
35 71422569324.1942
36 76526286649.8224
37 53006926046.8109
38 66880989508.9867
39 64560216623.2559
40 51814944973.1109
41 50149695825.5608
42 56449122113.8795
43 46210657615.119
44 40557757470.3059
45 38208194075.8606
46 38643678788.8573
47 36215736073.5292
48 36096597618.1036
49 33215035100.9787
50 30401737511.9298
};
\addlegendentry{0.6 Harvest}
\addplot [semithick, red]
table {%
1 252498370839.531
2 255334632438.21
3 251417470190.556
4 217561101305.665
5 214060551593.773
6 207836708523.208
7 206430659464.406
8 194850226599.675
9 213935079735.009
10 226626807704.356
11 208604017502.087
12 139700496151.202
13 156250849428.781
14 157488238064.521
15 180560739190.642
16 193386778065.623
17 133025878273.877
18 133840376871.275
19 143328741758.171
20 159426656861.165
21 98529913377.7673
22 113390930539.659
23 157136373223.356
24 86649532510.8304
25 105078305457.343
26 120576310711.644
27 79772188070.3076
28 125494260337.096
29 71372372468.5824
30 99055092032.8766
31 70942547037.5798
32 94403687922.6299
33 62473940664.3799
34 88323251627.5426
35 53557958227.269
36 74958694772.2215
37 67806486429.2567
38 48033653643.1081
39 63087273832.4431
40 59819036735.3773
41 43831298315.5432
42 48372189636.9686
43 47408381602.9117
44 44092434984.6695
45 42072116022.7969
46 38598086687.7576
47 35326985315.7519
48 36038692924.228
49 32748829759.7675
50 29075016667.2529
};
\addlegendentry{0.9 Harvest}
\addplot [semithick, blue]
table {%
1 169360794643.155
2 169360794643.155
3 169360794643.155
4 169360794643.155
5 169360794643.155
6 169360794643.155
7 169360794643.155
8 169360794643.155
9 169360794643.155
10 169360794643.155
11 169360794643.155
12 169360794643.155
13 169360794643.155
14 169360794643.155
15 169360794643.155
16 169360794643.155
17 169360794643.155
18 169360794643.155
19 169360794643.155
20 169360794643.155
21 169360794643.155
22 169360794643.155
23 169360794643.155
24 169360794643.155
25 169360794643.155
26 169360794643.155
27 169360794643.155
28 169360794643.155
29 169360794643.155
30 169360794643.155
31 169360794643.155
32 169360794643.155
33 169360794643.155
34 169360794643.155
35 169360794643.155
36 169360794643.155
37 169360794643.155
38 169360794643.155
39 169360794643.155
40 169360794643.155
41 169360794643.155
42 169360794643.155
43 169360794643.155
44 169360794643.155
45 169360794643.155
46 169360794643.155
47 169360794643.155
48 169360794643.155
49 169360794643.155
50 169360794643.155
};
\addlegendentry{Baseline}
\addplot [semithick, green01270]
table {%
1 176476240996.776
2 176476240996.776
3 176476240996.776
4 176476240996.776
5 176476240996.776
6 176476240996.776
7 176476240996.776
8 176476240996.776
9 176476240996.776
10 176476240996.776
11 176476240996.776
12 176476240996.776
13 176476240996.776
14 176476240996.776
15 176476240996.776
16 176476240996.776
17 176476240996.776
18 176476240996.776
19 176476240996.776
20 176476240996.776
21 176476240996.776
22 176476240996.776
23 176476240996.776
24 176476240996.776
25 176476240996.776
26 176476240996.776
27 176476240996.776
28 176476240996.776
29 176476240996.776
30 176476240996.776
31 176476240996.776
32 176476240996.776
33 176476240996.776
34 176476240996.776
35 176476240996.776
36 176476240996.776
37 176476240996.776
38 176476240996.776
39 176476240996.776
40 176476240996.776
41 176476240996.776
42 176476240996.776
43 176476240996.776
44 176476240996.776
45 176476240996.776
46 176476240996.776
47 176476240996.776
48 176476240996.776
49 176476240996.776
50 176476240996.776
};
\addlegendentry{Baseline}
\addplot [semithick, goldenrod1911910]
table {%
1 185639237871.023
2 185639237871.023
3 185639237871.023
4 185639237871.023
5 185639237871.023
6 185639237871.023
7 185639237871.023
8 185639237871.023
9 185639237871.023
10 185639237871.023
11 185639237871.023
12 185639237871.023
13 185639237871.023
14 185639237871.023
15 185639237871.023
16 185639237871.023
17 185639237871.023
18 185639237871.023
19 185639237871.023
20 185639237871.023
21 185639237871.023
22 185639237871.023
23 185639237871.023
24 185639237871.023
25 185639237871.023
26 185639237871.023
27 185639237871.023
28 185639237871.023
29 185639237871.023
30 185639237871.023
31 185639237871.023
32 185639237871.023
33 185639237871.023
34 185639237871.023
35 185639237871.023
36 185639237871.023
37 185639237871.023
38 185639237871.023
39 185639237871.023
40 185639237871.023
41 185639237871.023
42 185639237871.023
43 185639237871.023
44 185639237871.023
45 185639237871.023
46 185639237871.023
47 185639237871.023
48 185639237871.023
49 185639237871.023
50 185639237871.023
};
\addlegendentry{Baseline}
\addplot [semithick, red]
table {%
1 166703593272.714
2 166703593272.714
3 166703593272.714
4 166703593272.714
5 166703593272.714
6 166703593272.714
7 166703593272.714
8 166703593272.714
9 166703593272.714
10 166703593272.714
11 166703593272.714
12 166703593272.714
13 166703593272.714
14 166703593272.714
15 166703593272.714
16 166703593272.714
17 166703593272.714
18 166703593272.714
19 166703593272.714
20 166703593272.714
21 166703593272.714
22 166703593272.714
23 166703593272.714
24 166703593272.714
25 166703593272.714
26 166703593272.714
27 166703593272.714
28 166703593272.714
29 166703593272.714
30 166703593272.714
31 166703593272.714
32 166703593272.714
33 166703593272.714
34 166703593272.714
35 166703593272.714
36 166703593272.714
37 166703593272.714
38 166703593272.714
39 166703593272.714
40 166703593272.714
41 166703593272.714
42 166703593272.714
43 166703593272.714
44 166703593272.714
45 166703593272.714
46 166703593272.714
47 166703593272.714
48 166703593272.714
49 166703593272.714
50 166703593272.714
};
\addlegendentry{Baseline}
\end{axis}

\end{tikzpicture}

    \end{tabular}%
  \label{tab:addlabel}%
  \caption{Change curve of benefit and cutoff age with various wetness and wildfire threshold}
\end{figure}%

\subsubsection{Effects of Wood Price and Carbon Dioxide Emission Price}

As modeled, Net Primary Production (NPP) is essential to the cycling of the carbon, which is defined as the Gross Primary Production (GPP), that is, the overall amount of organic matter that all photosynthetic organisms can consolidate from environmental $CO_2$ minus the respired $CO_2$ from them when daylight available. Historically, to increase the amount of $NPP$ and presumably to decrease the value of $NEE$ - defined as the negative of $NPP$ - multiple nations have established a Kyoto Protocol, in which a global price for carbon is established based on country characteristics. We are interested in, therefore, implementing the utility function model by taking into account the different countries' forest coverage rate and past $CO_2$ emission.
We consider annual data of Country's Forest coverage area, Country's forest-contributing GDP, alongside Climate Factors including temperature and humidity from open data sources. We normalize these data by ranking these values through Z score distribution, with the formula being the following, and $\mu$ being the average and $\sigma$ being the standard deviation. 
\begin{equation}
\nonumber
    Z = (x-\mu)/\sigma
\end{equation}

\begin{figure}[H]
    \centering
    \includegraphics[width=0.4\textwidth]{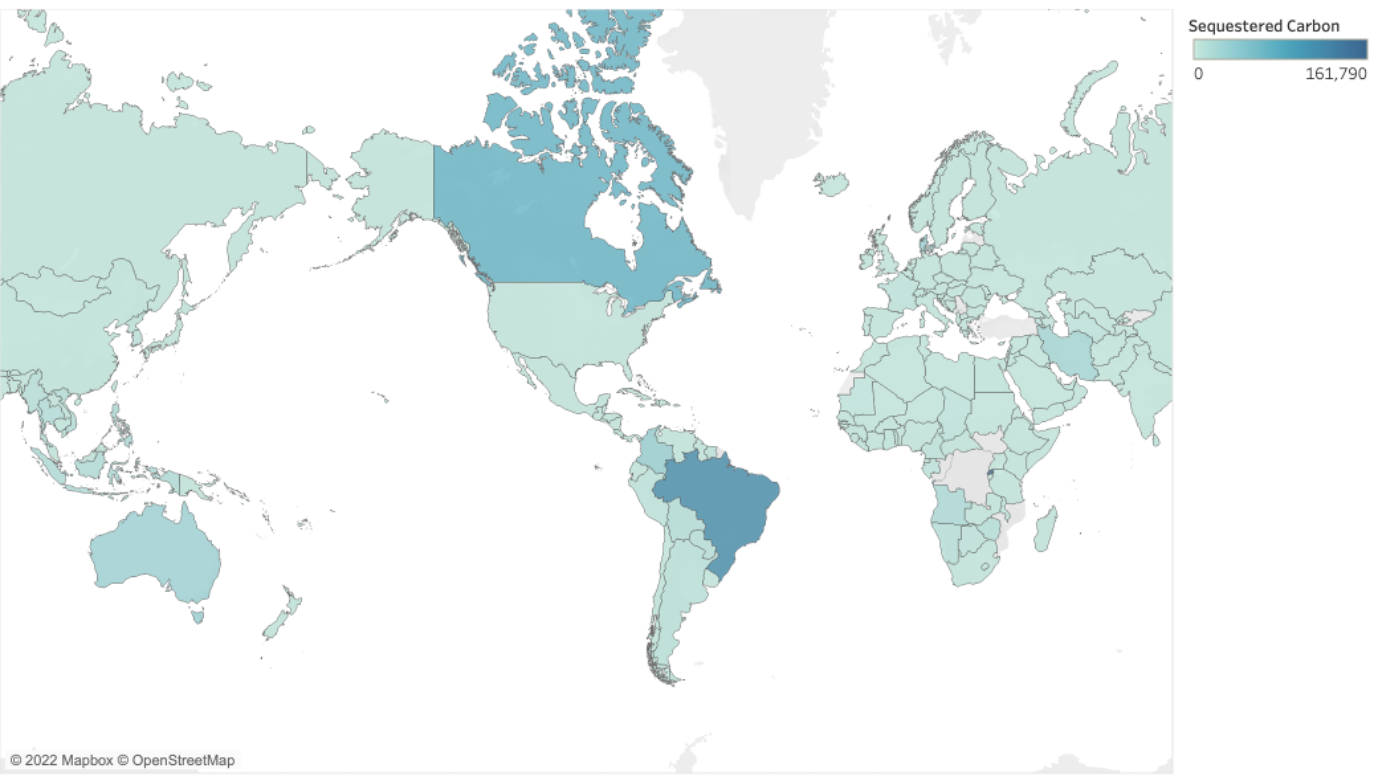}
    \caption{Geographical mapping of carbon sequestration model}
    \label{fig:my_label}
\end{figure}

\begin{figure}[H]
    \centering
    \includegraphics[width=0.4\textwidth]{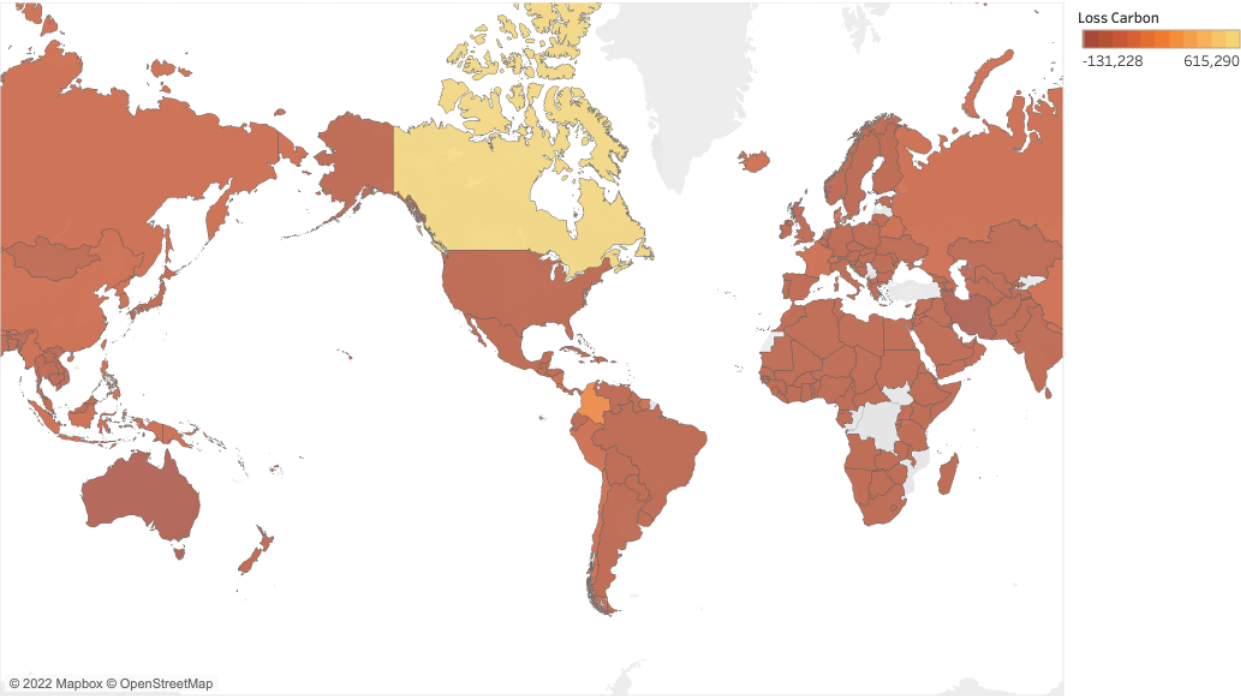}
    \caption{Geographical mapping of carbon loss}
    \label{fig:my_label}
\end{figure}

\section{Case Analysis}

We choose Georgia state to be our region of interest. As a Southeastern state, Georgia boasts itself with the most forest coverage in the United States$^{[7]}$. With the majority of the state in humid subtropical climate, Georgia becomes excellent in providing growth of forest region. Furthermore, Georgia’s climate has been inevitably warming due to anthropogenic activities. What the region may act to save the continuous increase of temperature while not exceeding the budget on managing the at-risk situation has become crucial.

Further, one of its key cities, the metropolitan downtown area, Atlanta, needs economic incentive for supplying the demand of local people. Therefore, the state becomes our exemplary model for analyzing the effectiveness of modified forest management strategies. We attempt to demonstrate the model's performance of carbon storage model and economic return model.

In the following sections, we zoom into Red Spruce as the sample species and use our model to help decide the status quo of forest conditions in Appalachian area and strategies of how to improve the forest management. We visualize the carbon storage along with the economic returns to illustrate the cost and benefit for further actions suggested by our modifications.

\subsection{Carbon Storage Benefits of Red Spruce in Appalachian Region}


Picea rubens, commonly known as Red Spruce, is a native species to eastern North America. More specifically, its distinct adaptation to the ecosystem of southern Appalachian spruce-fir forest. It is specifically adapted to high altitude, medium amount of moisture and grows on well-drained sand land. Its peculiarity of being perennial provides a good fit to the assumption of our model to predict the actions in the near hundred years. 

Such species has the average height of 5 to 54 meters and grow 10.6cm on a yearly basis. Its radius would therefore change accordingly. The average longevity range from 250 to more than 450 years. $^{[7]}$ This long time of growth results in a long term of sprout to mature stage when Red Spruce absorb the most carbon. We take data from literature, and estimate the function of Age and Carbon Dioxide Absorption to be (illustrated by the figure below):
\begin{equation}
\nonumber
    -0.008A^2+\ln(A)+8
\end{equation}

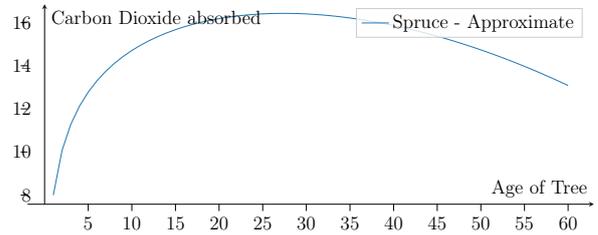
\begin{figure}[H]
  \center
    \begin{tabular}{rrrrrrrrrrrrrrr}
\begin{tikzpicture}[scale = 0.6]

\definecolor{darkgray176}{RGB}{176,176,176}
\definecolor{lightgray204}{RGB}{204,204,204}
\definecolor{steelblue31119180}{RGB}{31,119,180}

\begin{axis}[
legend cell align={left},
legend style={
  fill opacity=0.8,
  draw opacity=1,
  text opacity=1,
  anchor=north east,
  draw=lightgray204
},
tick align=outside,
tick pos=left,
x grid style={darkgray176},
xlabel={Age of Tree},
xmin=-1.95, xmax=62.95,
xtick style={color=black},
y grid style={darkgray176},
ylabel={Carbon Dioxide absorbed},
ymin=7.57642447009935, ymax=16.8510861279136,
ytick style={color=black}
]
\addplot [semithick, steelblue31119180]
table {%

1 7.998
2 10.0714415416798
3 11.2778368660043
4 12.1268830833597
5 12.7783137373023
6 13.3032784076842
7 13.7397304471659
8 14.1103246250395
9 14.4296737320087
10 14.7077552789821
11 14.9516858183951
12 15.166719949364
13 15.3568480723846
14 15.5251719888458
15 15.6741506033066
16 15.8057661667193
17 15.9216400321686
18 16.0231152736885
19 16.1113169374993
20 16.187196820662
21 16.2515673131703
22 16.3051273600749
23 16.3484826477874
24 16.3821614910438
25 16.4066274746046
26 16.4222896140644
27 16.429510598013
28 16.4286135305256
29 16.4198874899594
30 16.4035921449865
31 16.3799616134554
32 16.3492077083992
33 16.3115226843994
34 16.2670815738485
35 16.2160441844682
36 16.1585568153683
37 16.0947537379327
38 16.0247584791792
39 15.9486849383889
40 15.8666383623418
41 15.7787162001129
42 15.6850088548501
43 15.5856003470807
44 15.4805689017548
45 15.369987469311
46 15.2539241894673
47 15.1324428051302
48 15.0056030327237
49 14.8734608943319
50 14.7360690162844
51 14.593476898173
52 14.4457311557443
53 14.2928757406564
54 14.1349521396928
55 13.9719995556974
56 13.8040550722054
57 13.6311538035037
58 13.4533290316393
59 13.2706123317172
60 13.0830336866663

};
\addlegendentry{Spruce - Approximate}
\end{axis}

\end{tikzpicture}

    \end{tabular}%
  \label{tab:addlabel}%
  \caption{Red Spruce Tree Age (in decades) vs. $CO_2$ absorbed}
\end{figure}%

in which the visualization shows a prime age of carbon absorption at 19.75\% of the total living period before death. At the point shown, the ability of carbon storage peak for Red Spruce for a rather short time and soon decrease until it reach the stage of late maturity and death.

While the carbon storage dumped with aging, the direct logging of Red Spruce may be subject to insect parasitism, acid rains, and regional warming. Instead of cutting down directly the aged Red Spruce, we should approximate with the model to find the ones that are already severely damaged by insects and can barely benefit the surrounding natural environment.


\subsection{Economic Benefit of Red Spruce}

With mature technology of wood processing, Red Spruce plays a crucial role in the production of paper pulp, acoustic instrument, and even food productions. Its wide varieties of usage promise a logged Red Spruce just after its death would directly decrease the forest density, raise the related local economic income, and remove the carbon storage to allow more species in their prime age to attribute to carbon storage. 

With the current Carbon Volume and Price Ratio of 500\$ per tree, as well as the 20\$ per kilogram $CO_2$ of emitted based on Kyoto Protocol on Annex I countries, we recognize that carefully cutting down the aged Red Spruce which are impotent is critical in contributing in the natural may give more in the economic markets.

\subsection{Conclusion of Red Spruce Region}
Current Georgia management practices involve the use of controlled fire as a strategy. However, in such regard, carbon dioxide would be emitted from the burned trees, thereby creating a greater carbon source rather than flux. The Georgia Forestry Commission announces the probability of fire from a scale of 1 to 5, with the values closely mimick our simulation. Our strategy, in comparison with previous practices, aims at more specifically target at trees that need to be managed, while this may bring about an additional cost of human resources, the overall benefits are comparably huge. We take the above values into the mathematical equations and obtain the following strategies to stratify the forest environment. Additional data are obtained from the open data file from US Forest Service Department of Agriculture (2022), in which tree density is estimated based on available Growth-Removal-Mortality Model, which closely aligns with our computer simulation. Pandas module was imported in aiding the analysis of the problem.

Based on FIA data, our rough estimate concerning Georgia tree occupancy of the entire state is 15.4 billion. With a absorption rate of 21 kg / year, we have the total absorption without consideration of forest fire to be:
\begin{equation}
\nonumber
    21\times15.4\times 10^9 \times 10^-3 = 3.3\times10^8 \text{ton}
\end{equation}

We also obtain average forest fire affected acres from Fire and Smoke provided by the Augusta Chronicle, and assume all the carbon dioxide that the trees have absorbed over a lifetime was returned to the environment, and obtain the following two parts:

\begin{align}
    \text{Total } CO_2 &= \text{Wild} + \text{Human Induced} \nonumber \\
    &= 10\times \frac{458861}{60} \times 3.3\times (10^8) \nonumber \\
    &\quad \times \frac{1}{15.4\times10^9} \times 21 \times 100 \nonumber \\
    &= 344145 + 5601729 \nonumber \\
    &= 5.9\times10^6\text{ton} \nonumber
\end{align}

This amount of $CO_2$ loss is equivalent to, as in USD:
\begin{equation}
\nonumber
    (5.9\times10^6)\times 20 =  1.2\times10^7 \text{USD}
\end{equation}
Meanwhile, based on commodity price of carbon sequestered product, we estimate that with clear cut forests, the obtained carbon otherwise burned would yield a \textbf{gain} of the following economic benefits, equivalent as:
\begin{align}
    E &= \frac{458861}{60} \times 3.3\times (10^8) \nonumber \\
    &\quad \times \frac{1}{15.4\times10^9} \times 21 \times 100 \times 20 \nonumber \\
    &= 6.9\times 10^6 \text{USD} \nonumber
\end{align}

Given these data above, we have therefore quantitatively demonstrated an increase in the overall benefits as far as current tree management strategies are concerned. 

\section{Discussion: Strengths and Weaknesses}
\subsection{Strengths}

\begin{itemize}
\item Considering various indicators in environmental aspects, we also combined possible factors of economic returns to make a more comprehensive and objective model.

\item The simulation process may be replicated while changing the key conditions of average longevity of a specific forest, making our model effective and valuable under different environmental settings.

\item The model of the carbon storage considers the influence of humidity, tree volume, tree growth rate, and the process of photosynthesis separately to fully identify the key linkages of the carbon cycling system.
\end{itemize}

\subsection{Weaknesses}

\begin{itemize}
    \item With the more drastic climate changes which result in rise of sea level, acid rains, and extreme weathers, the benefit of our model won’t last forever. Upgrades are needed for the new dominating factors to adjust continuously for even longer term of development. 
    
    \item Some data is difficult to obtain and is dependent on rules of thumb encountered in biology, which may therefore cause sample errors.
\end{itemize}

\section{Conclusion and Results}{Logging as a Part of Long-term Forest Management}

To address the ongoing discussion on the forest condition related to climate change, we conclude that it is unacceptable to pursue zero-logging for the so-called "eco-consciousness", as it may easily turn into extremism. Logging, when done appropriately, can prevent unfavorable situations, such as wildfires, which are caused by high density of trees in the forest. Hence, we investigated a considerable number of variables and found the relationship between proper logging and carbon dioxide.
\\

Recent environmental protection strategies tend to favor the diminishing human activities on the natural environment as much as possible. In terms of active pollution and over-harvesting, such activities are not commendable, but we should also realize that human intervention can sometimes play a positive role. When done correctly, human intervention can help stabilize ecosystems. Providing underdeveloped areas with economic support to harvest carbon products—through creating tourism opportunities for rural areas—can also benefit communities and the environment at large. All things considered, an arbitrary closure of all-natural forests toward human beings is an impractical approach, and it is not economically feasible in the long run.
\\

Species density plays a substantial role in advocating for proper forest management strategies. As known in previous literature, drought occurs cyclically. When using computations to model tree growth in arid areas within a simulated environment, data reveals several potential factors of harm: a high likelihood of forest fires, consequently loss of carbon storage, an imbalance between carbon release and absorption, and further diminishing of keystone species with high volume of carbon storage (even in droughts). As a result, the anticipated species diversity would remain low despite the efforts of maintaining the secluded environment. In contrast, when implementing an age-based approach towards logging of trees, though the area suffers from a lack of water in arid times, this method effectively protects the trees against the possibility of forest fire. Furthermore, the obtained logs help ensure a decent income for the local economy and uplifts the community with natural resources, thereby balancing the loss of carbon in the process of sequestration.
\\

In most situations, the area does not suffer from year-round drought, the surrounding industry does not depend actively on harvesting wood, nor does local water loss and soil erosion greatly rely on the number of trees. Thus, an appropriate amount of tree logging would protect the carbon storage in forest while yielding sustainable economic return.
\\

Humans, by our presence upon this Earth, is already impacting every inch of the environment, be it directly or indirectly. And it is time for us to collectively be responsible for the fact that we owe it both to the Earth and to ourselves to draw up our next steps with sustainable development. Instead of blindly taking extreme approaches due to shallow look-ups, we should rather remain a humble and careful approach with thorough researches, such that humans can maintain harmony with the Earth and the countless species who also call this planet "home."



\end{multicols}

\newpage
\includepdf[pages=1,pagecommand={}]{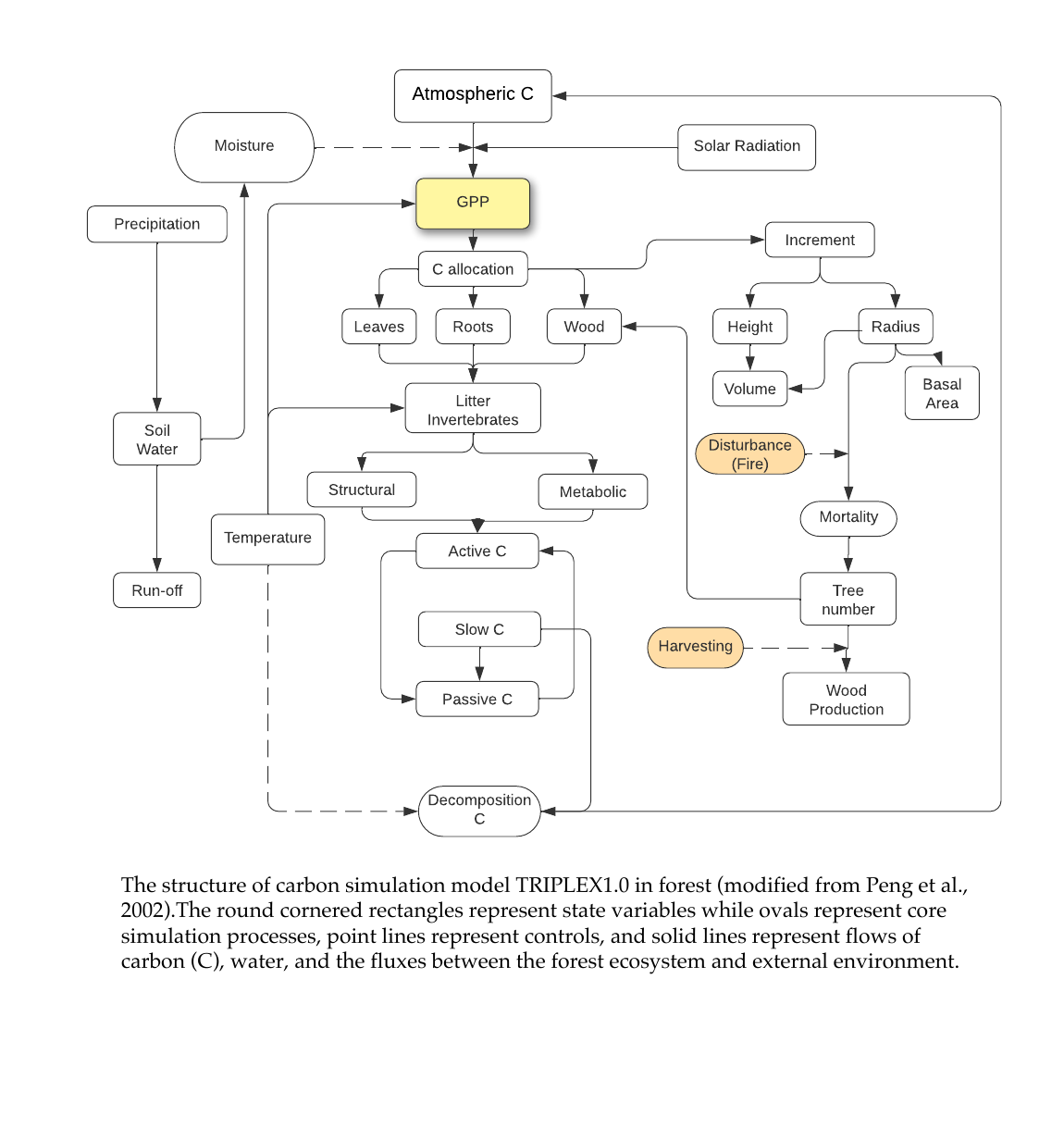}
\includepdf[pages=1,pagecommand={}]{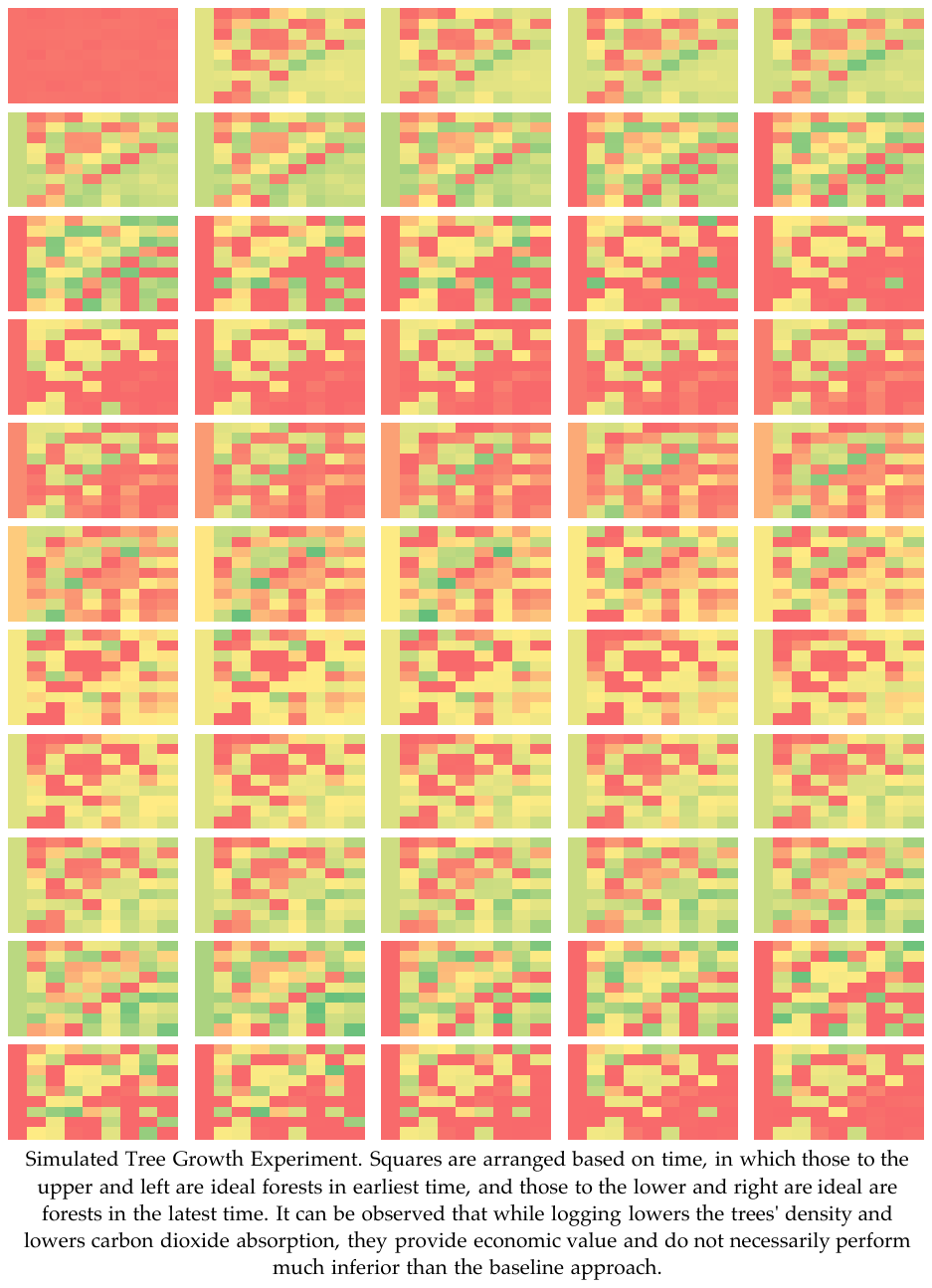}

\newpage

\begin{table}[htbp]
\begin{center}
\caption{Notations}
\begin{tabular}{cl}
	\toprule
	\multicolumn{1}{m{2cm}}{\centering Symbol}
	&\multicolumn{1}{m{10cm}}{\centering Definition}\\
	\midrule
    $C_{tot}$&represents the overall carbon storage\\
	$t$&represents the time passed since the origin, set the original time at $t(0)$ \\
	$T$&represents a fluctuating temperature of the given environment\\
	$L(t, T)$&represents the overall humidity of the modeled natural \\
	$\quad$ & due to time and temperature of surrounding environment\\
	$h$&represents the height of the photosynthetic organism\\
	$rad$&represents the radius of the tree\\ 
	$s$&represents the amount of chopped down trees\\
	$S$&represents carbon stored per unit volume of stem\\
	$A$&represents the age of the tree\\
	$\lambda$&represents the geometric factor\\
	$m$&represents the mass of the tree at time t\\
	$\rho$&represents the average density of carbon storage in each tree\\
	$\psi$&represents the water potential\\
	$E$&represents the economic return due to carbon storage\\
	$P_1$&represents the cost of physical and human capital\\
	$P_2$&represents the economic expenses on getting economic return\\
	$\tau$&represents the ratio of the number of drought resistant\\
	$\quad$&Keystone Species over the number of non-resistant common species\\
	$I$&represents the total budget may be put in controlling carbon storage\\
	$Simp$&represents the Inverse Simpson index which indicate \\
	$\quad$&higher species diversity and consequently a higher possibility of carbon fixation\\
	\bottomrule
\end{tabular}\label{tb:notation}
\end{center}
\end{table}

\end{document}